\documentclass[twocolumn]{aastex63}

\usepackage{float}

\usepackage{changepage}
\usepackage[section]{placeins}

\usepackage{acronym}
\usepackage{hyperref}
\usepackage{amsmath,amsfonts,amssymb}
\usepackage{mathrsfs}
\usepackage{mathtools}
\usepackage[utf8]{inputenc}

\date{Received August 14, 2026}

\shorttitle{Birth Function of Neutron Stars}

\begin{document}

\title{Comparison of the Neutron Star Birth Function from Core-Collapse Supernova Simulations with Observations}
\author[0009-0005-8878-8756]{Waly M Z Karim}
\email{walym.zkarim64@gmail.com,aburrows@princeton.edu}
\affiliation{Department of Physics, University of Rochester, Rochester, NY, 14627, USA}
\author[0000-0002-3099-5024]{Adam Burrows}
\affiliation{Department of Astrophysical Sciences, Princeton University, NJ 08544, USA}
\author[0000-0003-1938-9282]{David Vartanyan}
\affiliation{College of Science, University of Idaho, 875 Perimeter Drive, Moscow, ID 83844}
\author{Anthony Fisher}
\affiliation{Department of Physics, Princeton University, NJ 08544, USA}
\author{James Burry}
\affiliation{Department of Physics, Princeton University, NJ 08544, USA}

\begin{abstract}
Using the mapping between progenitor core structure and the gravitational mass of neutron stars derived from sophisticated 3D supernova simulations, we construct the theoretical mass distribution of neutron stars at birth and investigate how it might comport with neutron star mass function measurements. In the process, we explore the effects of islands of black hole formation and subsequent mass accretion. We show that supernova theory can explain the observed neutron star mass function from its lower-mass peak near $\sim$1.35 $M_{\odot}$ to its higher-mass tail.   Moreover, we suggest, but do not prove, that the lower predicted kick speeds expected during the birth of lower-mass neutron stars and the higher expected kick speeds expected on average for higher mass neutron stars may together sculpt the observed mass function in desired directions. The upshot of all these influences is to imprint upon the measured neutron star mass function features that reflect the varied physics of both neutron star origins and the neutron-star/black-hole dichotomy. Very approximately, we derive a  black hole birth fraction of $\sim$19.6\%. In summary, we suggest that supernova theory can
now be used to explain, however provisionally, various
measured attributes of the population of compact objects and that an era of productive engagement between
supernova theory and observation is at hand.
 \end{abstract}

\keywords{stars - supernovae - general }

\section{Introduction}


After many decades of toil and progress, core-collapse supernova theory is now witnessing explosions regularly,  naturally, and without artifice. This is due in part to the incorporation of increasing degrees of complexity in 1) numerical technique, 2) neutrino transport, and 3) neutrino-matter interaction physics, and to the size and power of the modern international computational infrastructure. Reviews and representative papers of note together catalog and explain the current perspective and theoretical terrain \citep{janka2012,janka2025,lentz:15,burrows2013,burrows_2020,wang2022,Burrows2024,2021Natur.589...29B,muller2017,Muller2019,2020ApJ...896..102K,2018MNRAS.477L..80K,roberts:16}.

However, there is still much to be done in detail to understand 1) the role of initial progenitor structure, 2) the distribution of outcomes due to the chaotic nature of the flow, and 3) the reasons for remaining variations from group to group. Nevertheless, sophisticated 3D supernova theory is now emerging to be a predictive enterprise, with results connecting to measurements of supernova light curves \citep{vartanyan_breakout_2025,vartanyan2026_Ni}, pulsar proper motions \citep{2017ApJ...842...13W,janka_kicks_2024,coleman,spin2023}, nucleosynthesis \citep{wanajo2018,sandoval2021,sieverding2023,wang_nucleo_2024,wind,janka2025}, explosion energies \citep{radice2017b,Muller2019,stockinger2020,wang_low_2024,Burrows2024}, and debris morphologies and ejecta element distributions in space and velocity \citep{2021MNRAS.502.3264G,sandoval2021,vartanyan_breakout_2025,vartanyan2026_Ni,orlando2025_87A,orlando2025_CasA}. One can't yet trust the predictions and results in detail, but the coupling between theory and observation is palpably improving. In particular, and what motivates this paper is the suggestion in \citet{Burrows2024} that there could be a monotonic relationship between the gravitational mass ($M_{\mathrm{birth}}$) of a neutron star birthed in a core-collapse supernova explosion and the final core structure of the progenitor massive star,
in particular its ``compactness" \citep{Oconnor2011} (see \S\ref{method} and Table \ref{table1_new}). With a mapping between compactness and residual neutron star mass, a mapping between compactness and ZAMS\footnote{zero-age-main-sequence }mass, and a ZAMS-mass mass function (such as the \citet{salpeter} distribution), one can derive a mass function of neutron stars at birth. There is a long literature on the gravitational masses of neutron stars in binaries (see \S\ref{measured}). And there is a long literature on inferences of the underlying neutron star mass function, with tentative explorations into the effects of subsequent mass accretion and of selection bias. The latter might be quite severe, but is not fully characterized. Nevertheless, we feel the time is ripe for a range of predictions for the underlying birth mass function in the context of modern, however imperfect, supernova theory. This is the motivation for this paper.

In \S\ref{measured_1}, we summarize the current knowledge of the measured mass distributions of neutron stars and what has to date been inferred. In \S\ref{method}, we present our method for approaching the calculation of a birth mass function of neutron stars, with asides on our simple provisional handling of the subset of birthed stellar-mass black holes and on the corrections due to subsequent mass accretion from a companion. Then, in \S\ref{results}, we calculate our default birth mass function, as well as mass functions using various variations on our default assumptions and inputs. In \S\ref{conclusion}, we reflect on the results obtained and speculate on the road forward to tighten up our predictions and to foster improvements in the analysis of the populations of measured neutron stars and its important connection with supernova theory.

\section{Measured and Inferred Neutron Star Mass Functions}
\label{measured_1}

\subsection{Measured Masses}
\label{measured}

\begin{figure*}[htbp]
        \includegraphics[width=\columnwidth]{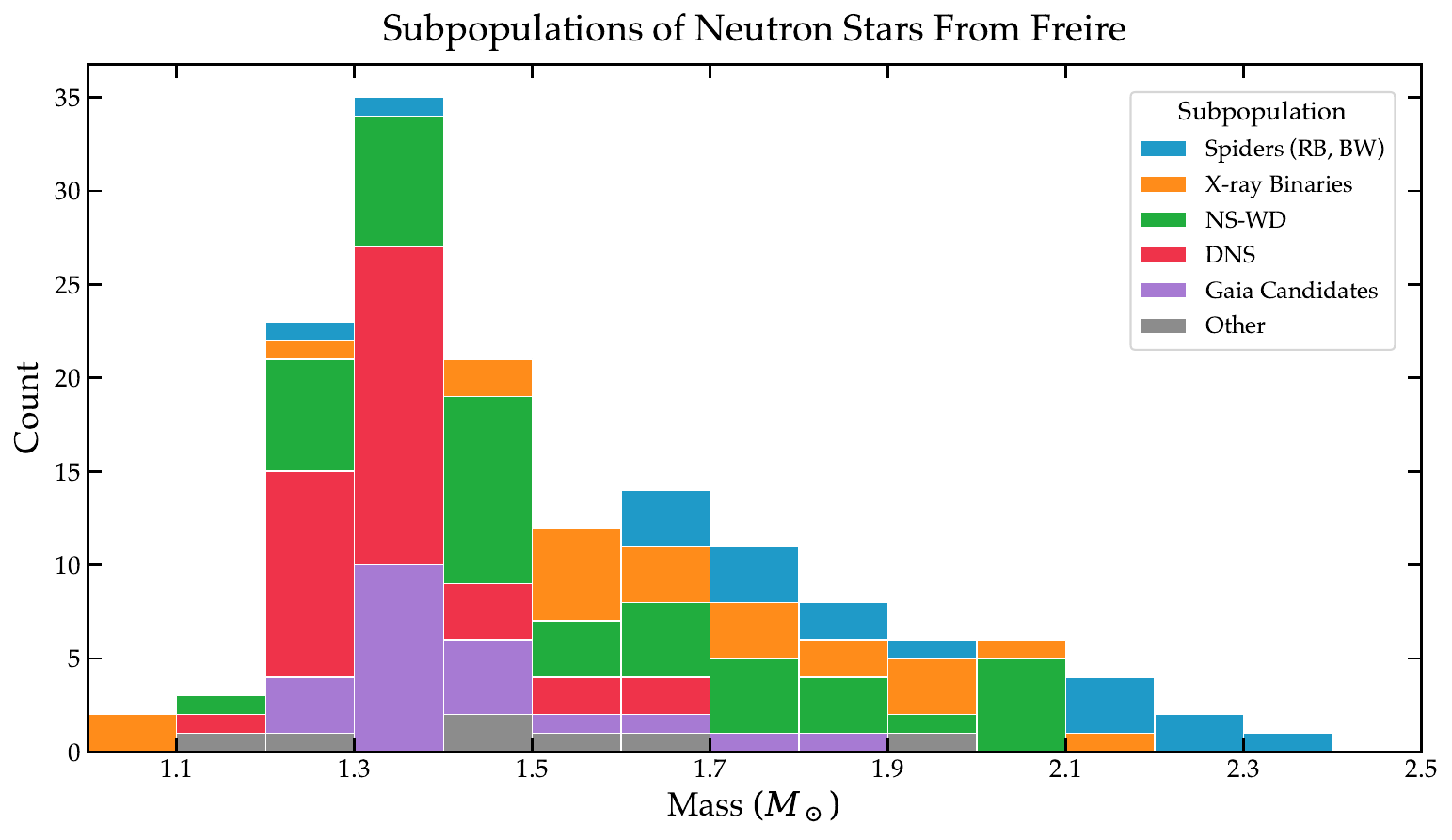}
        \includegraphics[width=\columnwidth]{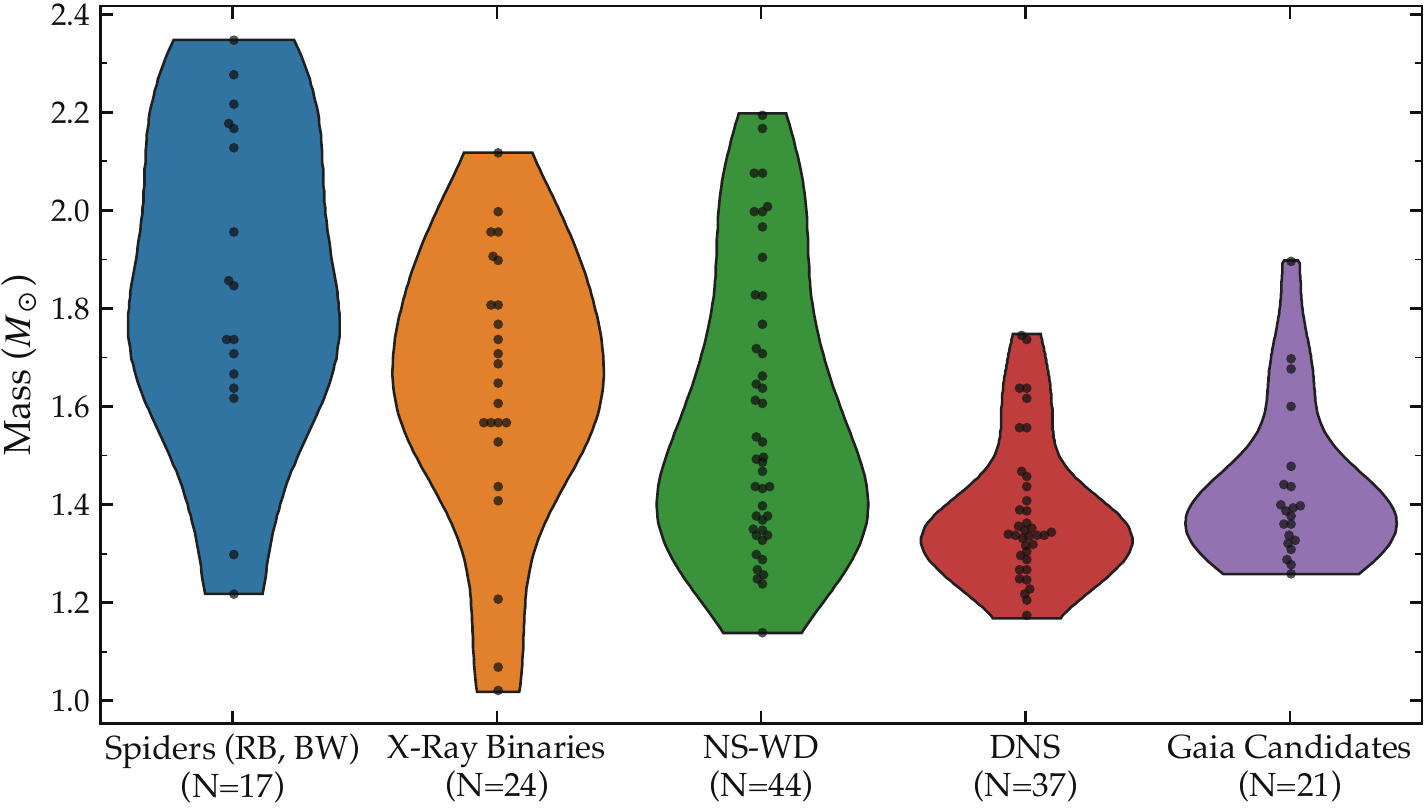}
    \caption{\small Observed neutron star binary data for objects between 1.0 and 2.4 $M_{\odot}$, grouped by subpopulation type. Left: Stacked histogram with bars colored according to the proportion accounted for by each subpopulation. Right: Violin plot showing the distribution per subpopulation. Observed data points are shown at the center of each violin; the widths correspond to a summed Gaussian kernel over the data points. The subpopulations are shown on the x-axis. Note that the NS-NS subpopulation is weighted at the low-mass end.  This feature is explained in \citet{spin2023} as a result of the lower kick speeds at birth expected when the lowest mass neutron stars are born, thereby slightly favoring the survival of the binary. Hence, this pattern is deemed by \citet{spin2023} a selection effect. See text for further discussion. Data and references are tabulated in the Appendix.}
    \label{fig:stacked_utd_hist}
\end{figure*}

There is a rich literature on attempts to constrain the neutron star mass distribution function. All measured masses are for neutron stars in binaries, so there is an overarching selection bias that is generally not well characterized. Nevertheless, several studies have applied Bayesian statistical methods to neutron star mass observations and inferred distribution functions~\citep{Zhang2011,ozel2012,antoniadis,alsing,rocha2019,fan2023,fan2024,zwack2025}. Results have varied significantly depending on the observational sample considered. To date, there is little consensus about the mass function of neutron stars at birth, in part due to the remaining ambiguities due to selection bias and the magnitude of subsequent accretion from their companions. The number of neutron star mass estimates used in these studies has increased over time to the current dataset of 148 objects, shown in the left panel of Figure \ref{fig:stacked_utd_hist} collectively and in Table \ref{tab:unified_ns_masses}.

We divide neutron stars in binaries into five main subpopulations: Spiders, X-ray Binaries, Double Neutron Star (DNS) systems, Neutron Star-White Dwarf (NS-WD) Systems, and Gaia Candidates. For quick insight into their different observed distributions, the right panel of Figure \ref{fig:stacked_utd_hist} separates them into their respective violin plots. Mass determinations are also available for a small number of Neutron Star-Main Sequence (NS-MS) and Neutron Star-Black Hole Systems (NS-BH), but one generally excludes these objects from a mass function determination because the sample sizes for these cohorts are too small to be useful.

Spider systems are binaries in which a millisecond pulsar tightly orbits a lower mass companion that is a non-degenerate or semi-degenerate star (\cite{blanchard2025}). Spiders are divided into Black Widows, which have very low-mass companions, and Redbacks, which have higher-mass companions (\cite{You2025}; \cite{fan2024}). 

X-ray Binaries are systems in which the neutron star accretes material from its companion star. Systems with companions more massive than the primary are known as high-mass X-ray binaries (HMXBs) and systems with companions less massive than the primary are known as low-mass X-ray binaries (LMXBs) (\cite{ozel2012}; \cite{fornasini2023}; \cite{Bahramian_2023}).

\begin{figure}[htbp]
    \centering
        \centering
        \includegraphics[width=\linewidth]{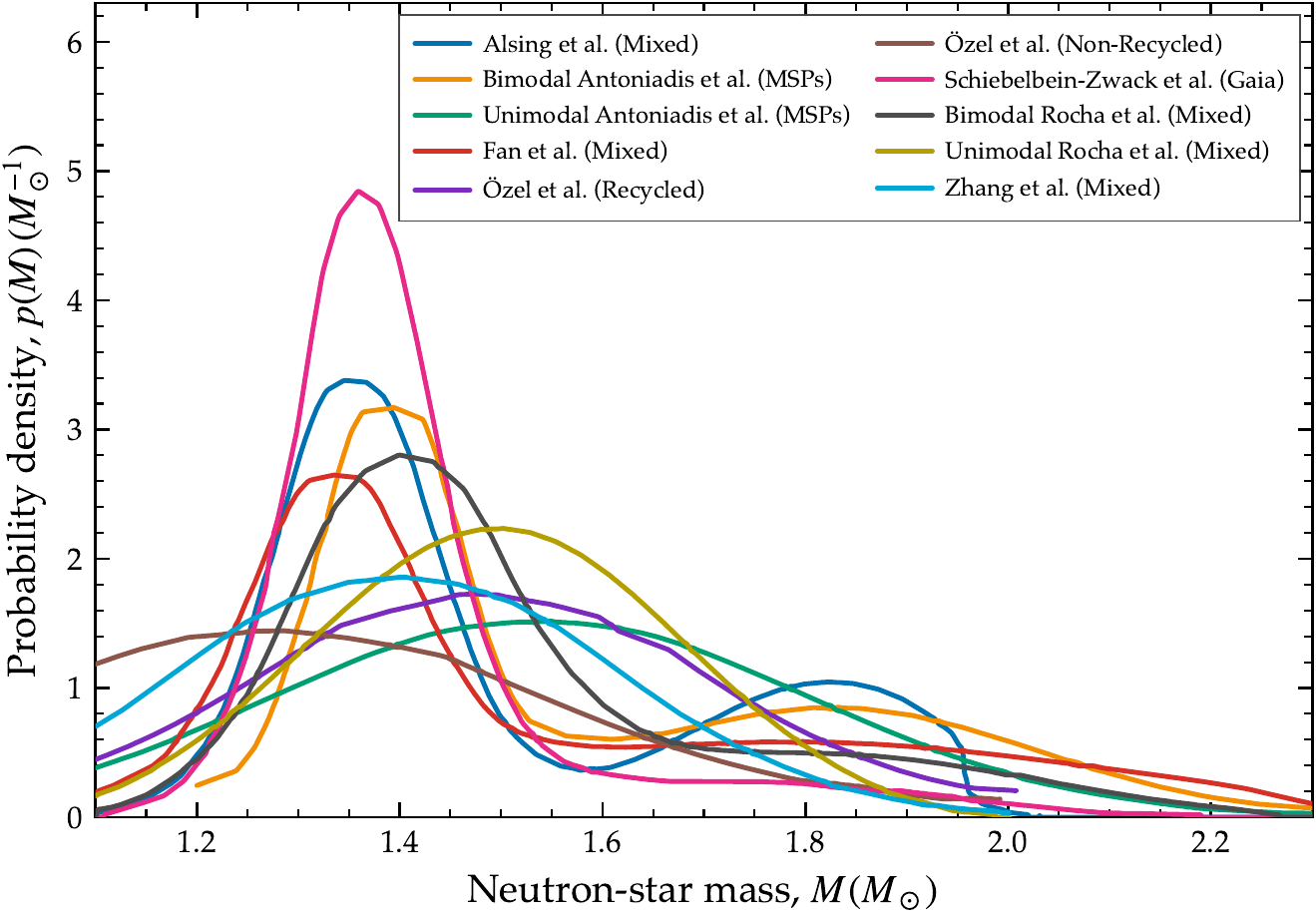}
    \caption{\small Inferred neutron star mass distributions from the literature, including both unimodal and bimodal Gaussian models. Sources are \cite{Zhang2011}, \cite{ozel2012}, \cite{antoniadis}, \cite{alsing}, \cite{rocha2019}, \cite{fan2024}, \cite{das2024}, and \cite{zwack2025}. Legend labels indicate the neutron star subpopulation used in each study; Antoniadis et al.\ and Rocha et al.\ each appear twice because both a unimodal and bimodal fit were derived for the full population.}
    \label{fig:theory_combined}
\end{figure}

DNS and NS-WD systems are those in which a neutron star is in orbit with either another neutron star or a white dwarf, respectively. The new Gaia observations are not yet well classified; recent studies (\cite{Badry2024Gaia,zwack2025,2026ApJ...998L..14S,2026arXiv261000454S}) found that their mass distribution is similar to that of Double Neutron Stars (light pink line in Figure \ref{fig:theory_combined}), but we consider them a separate subpopulation here. \citet{Badry2024Gaia} reported \(21\) astrometric binaries containing solar-type stars and dark companions with masses near \(1.4\,M_\odot\), selected from Gaia DR3 astrometric binary solutions and followed up with multi-epoch radial velocities over about \(700\) days to validate the astrometric orbits and tighten measurements of the companion masses. These authors selected systems in which the luminous star is on the main sequence and the dynamically implied dark companion mass is above \(1.25\,M_\odot\). Moreover, they found several companions unambiguously above the Chandrasekhar limit\footnote{These authors erroneously assume that the relevant Chandrasekhar mass is $\sim$1.4 $M_{\odot}$.  In fact, massive star evolution yields cores that collapse by the Chandrasekhar instability with masses between $\sim$1.3 and $\sim$2.0 $M_{\odot}$ due to the effects of thermal pressure support and variations in the electron fraction ($Y_e$). This fact will be germane in what follows in this paper.}, while the rest lie between \(1.25\) and \(1.4\,M_\odot\) \citep{Badry2024Gaia}. The Gaia sample is subject to some selection bias \citep{2026arXiv261000454S}: the candidates are mainly solar-type stars on the main sequence, the orbital periods are mostly \(100\text{--}1000\) days, the astrometric solutions are sensitive to orbital periods near the Gaia observing baseline, and the candidate set depends on follow-up radial velocities to reject spurious astrometric solutions and confirm the dark-companion masses \citep{Badry2024Gaia,2026arXiv261000454S}.

\subsection{Inferred Mass Functions}
\label{inferred}

 \citet{fin1994} was one of the first to infer the range of neutron star masses. Finn's study considered eight neutron star masses from DNS systems and found a range of 1.3 to 1.6 $M_{\odot}$. A larger sample of 19 neutron stars was then used by \cite{Thorsett1999} to derive a mean peak at 1.35 $M_{\odot}$ with a dispersion of 0.04 $M_{\odot}$.

Since then, more neutron star mass estimates have become available. \cite{Zhang2011} considered 61 total mass estimates in a sample of 18 DNS systems, 26 NS-WD systems, 3 NS-MS systems, 13 X-ray binaries, and one undetermined system (J1753-2240). They fitted their full `mixed' sample using a Gaussian curve centered at 1.46 $\pm$ 0.18 $M_{\odot}$. Their inferred mass function, along with those derived by others discussed below, is plotted in  Figure \ref{fig:theory_combined}.

\cite{ozel2012} considered a sample of 41 binary neutron stars from 5 DNS systems, 20 NS-WD systems, and 6 X-Ray Pulsar systems.  They then further divided these into likely recycled and non-recycled categories (more on this later) in an attempt to distinguish between sources with distinct evolutionary and accretion pathways. Their method produced unimodal mass distributions. DNS systems yielded a peak at 1.33 $M_{\odot}$ with a dispersion of 0.05 $M_{\odot}$. Non-recycled neutron stars peaked at 1.28 $M_{\odot}$ with a dispersion of 0.24 $M_{\odot}$ (brown line in Figure \ref{fig:theory_combined}). Recycled neutron stars had a higher peak of 1.48 $M_{\odot}$ with a dispersion of 0.2 $M_{\odot}$ (purple line in Figure \ref{fig:theory_combined}).

\cite{antoniadis} performed a similar analysis on a smaller sample of 32 millisecond pulsar masses (MSPs), comprised of 18 MSPs with precisely determined masses, 10 MSPs with constraints only on total mass, and 4 systems with constraints only on the mass ratio, $q$. Their study performed theoretical fits for both unimodal and bimodal models, finding evidence for a bimodal distribution with a first peak at $1.393^{+0.031}_{-0.029}$ $M_{\odot}$ and a second peak at  $1.807^{+0.081}_{-0.132}$ $M_{\odot}$ (orange line in Figure \ref{fig:theory_combined}). However, they did not reject the possibility of a skewed unimodal mass distribution (green line in Figure \ref{fig:theory_combined}). This study also addressed the question of the maximum mass and found a lower limit of 2.018 $M_{\odot}$ with a 98\% confidence level.

\cite{alsing} then compiled a larger population of 74 neutron star systems for which either more than two post-Keplerian parameters or X-ray/optical mass measurements were available. Unlike \cite{ozel2012} and \cite{antoniadis}, they considered the entire population together (`mixed' in Figure \ref{fig:theory_combined}) using a ``flexible Gaussian mixture model" designed to naturally identify subpopulations and allow for multiple modes which might result from distinct formation channels. They found evidence for a bimodal distribution with peaks at approximately 1.35 $M_{\odot}$ and 1.85 $M_{\odot}$ (blue line in Figure \ref{fig:theory_combined}), as well as a maximum cut-off mass in the interval [2.0, 2.2] $M_{\odot}$ with 68\% confidence level, and the interval [2.0, 2.6] $M_{\odot}$ with 90\% confidence level. These results were consistent with those obtained by \cite{antoniadis}.

\cite{farrow2019} built upon the study by \cite{ozel2012} following the ``breakthrough discovery" of the GW170817 system. They considered only those systems for which the superior radio pulsar timing observations were available, which resulted in a limited, but potentially more accurate, dataset of 17 DNS systems. Their Bayesian statistical approach produced evidence for a bimodal mass distribution of presumably recycled neutron stars and a unimodal distribution for non-recycled neutron stars. This result contradicted the unimodal distributions found in \cite{ozel2012}, but with the caveat that the sample size of 17 used by \citet{farrow2019} was insufficient to draw meaningful conclusions. \cite{farrow2019} estimated that 20 DNS systems would be required to confidently determine whether presumably recycled and non-recycled neutron stars share a common distribution. They suggested that approximately 60 DNS systems would be required to establish the shape of these distributions.

In the same year, \cite{rocha2019} performed two frequentist statistical tests (Kolmogorov-Smirnov and Anderson-Darling) on a sample of 73 neutron star masses. This sample consisted of 18 X-ray binaries, 16 DNS, 36 NS-WD binaries, and 3 NS-MS binaries. Their goal was to clarify whether observed objects grouped together in previous work did indeed belong to the same populations. A unimodal, bimodal, and multimodal mass distribution function produced using peak and dispersion values calculated from this dataset were each tested for their fit to the sample of 73 neutron stars. The frequentist tests rejected the possibility of a unimodal distribution (gray line in Figure \ref{fig:theory_combined}) and they concluded in favor of either a bimodal (light green line in Figure \ref{fig:theory_combined}) or multimodal profile. A common thread across these studies is that the observed sample contains a low-mass concentration near \(1.25\text{--}1.4\,M_\odot\), a broader or heavier (presumably) recycled component at larger masses, and an upper boundary whose interpretation is entangled with both the maximum mass issue and post-birth evolution \citep{Zhang2011,Ozel2016,antoniadis,alsing,You2025}. However, there is also an emerging consensus among observers that subsequent accretion might not be enough to explain the higher-mass tail, and some neutron stars might be born massive. As we discussed above, \citet{antoniadis} was one of the initial studies to point this out by finding a bimodal neutron star mass distribution, with a heavier peak at $1.807 M_\odot$. Other studies that also suggested bimodality in the birth mass distribution, and, hence, supported the claim that neutron stars can be born massive include ~\citet{alsing} and \citet{You2025}.

Subsequently, improvements in measurement and consequent discoveries of higher mass examples have prompted more detailed studies into the maximum cut-off mass for neutron stars. For example, \cite{2024PhRvD.109d3052F} considered a large sample of 136 neutron star masses comprised of 23 DNS systems, 68 NS-WD systems, and 23 X-ray binary systems and found a cut-off maximum of $2.25^{+0.08}_{-0.07} $ $M_{\odot}$ with a 68.3\% confidence level in a two-component Gaussian mixture distribution (red line in Figure \ref{fig:theory_combined}). As a result, they argued that the 2.5-3 $M_{\odot}$ compact objects observed by gravitational-wave detectors were unlikely to be neutron stars, instead low-mass black holes.

It is clear from this previous literature that there has been a presumption that the observed tail at higher masses is likely the result of subsequent mass accretion and that the peak near $\sim$1.4 $M_{\odot}$ is the actual birth mass tied to the presumed textbook Chandrasekhar mass.  However, as a result of corrections in the core mass at instability due to thermal pressure support at the final stages of progenitor star evolution \citep{swbj16,woosley2002}, theory clearly indicates that the effective Chandrasekhar core that collapses ranges in mass much more broadly from $\sim$1.3 to $\sim$2.0 $M_{\odot}$, depending predominantly upon the progenitor mass. Note, however, that though the final gravitational mass of the residual neutron star depends upon this collapsing core, it is modified by the dynamics of the supernova explosion and subsequent mass infall, corrected by the total neutrino mass-energy loss, which itself is a function of the nuclear equation of state (still not definitely constrained). The upshot is the expectation that supernovae leave a broad spectrum of neutron star gravitational masses and this is what recent supernova theory finds \citep{Burrows2024}. It is this finding that in part motivates this paper.  We ask the question: what now, in light of recent state-of-the-art 3D supernova simulations that predict a spectrum of birth masses, is the expected theoretical birth mass function of neutron stars? In this paper, we make the first attempt to combine the mappings of progenitor ZAMS mass to the ``compactness index" of core structure and of core structure to neutron star birth mass derived from supernova theory with the mass function of massive stars to derive a theoretical neutron star mass function. In addition, we explore the potential effects of subsequent mass accretion and of black hole formation on the neutron star mass function. 

Note that for core-compactness to neutron star birth mass mapping, we base our analysis on the results from~\citep{Burrows2024}. In the existing literature, there is a dearth of 3D radiation-hydrodynamic models of the CCSN phenomenon to late post-bounce times that provides a comprehensive compactness/progenitor to neutron star residue mass mapping. Other remnant-mass prescriptions do exist, including the probabilistic semi-analytic model of \cite{2020MNRAS.499.3214M} and the spherically symmetric ``1D+" calculations of \citet{Luca2024}, the latter providing a crude compactness versus birth mass fit. However, these approaches originate from simple prescriptions and they represent the multi-dimensional post-shock context, to the degree they do, only very approximately; they are not the product of detailed 3D simulations. In particular, in the 1D+ model in \citet{Luca2024}, the accretion ceases once the explosion sets in, whereas in fully 3D simulations accretion can continue for an extended period. We note also that although other groups have employed 3D simulations and have found correlations between core structure and neutron star birth mass~\citep[e.g.,][]{2025MNRAS.536..280N}, they have been evolved for a substantially shorter post-bounce time and do not explore a full progenitor mass range.

\section{Method}
\label{method}

\begin{figure}[htbp]
    \centering
    \includegraphics[width=\linewidth]{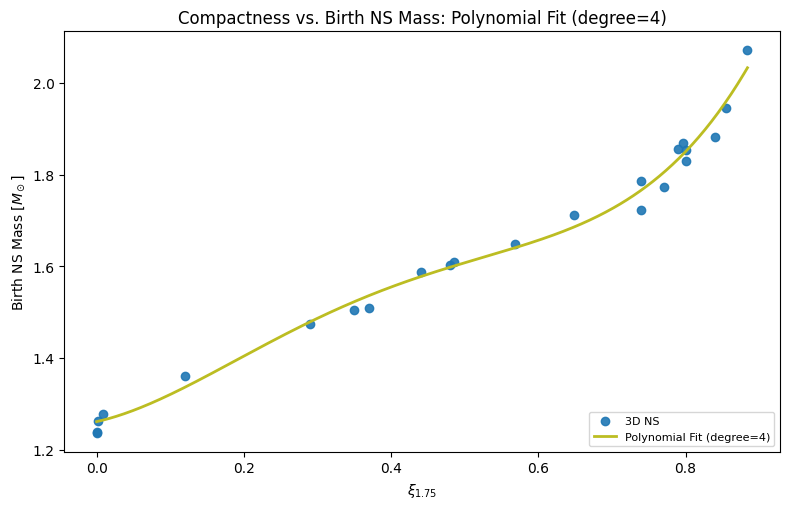}
    \caption{A representative polynomial fit for the compactness-gravitational mass ($M_{\mathrm{birth}}$, birth mass here) mapping. While linear, quadratic, and even cubic models were considered, the polynomial fit's curvature best captures the variation in the mass-compactness mapping. The data points for this mapping can be found in Table \, \ref{table1_new}. The gravitational mass is in units of \(M_\odot\).}
    \label{fig:polyfit}
\end{figure}

\citet{Burrows2024} reported that 3D supernova simulations carried to late post-bounce times suggested that neutron-star gravitational birth masses are roughly monotonic functions of compactness \citep{Oconnor2011}. The compactness parameter is defined by

\begin{equation}
\xi_M = \frac{M/M_\odot}{(R(M)/1000\, \mathrm{km})}\, 
\label{comp}
\end{equation}
where \(M\) is the enclosed presupernova Lagrangian mass coordinate and \(R(M)\) is the radius enclosing that mass coordinate at the time of collapse. Our analyses evaluate this parameter at \(M=1.75\, M_\odot\), and, thus, the parameter used is \(\xi_{1.75}\). The results of \citet{Burrows2024} provide a direct mapping between the compactness \(\xi_{1.75}\) of the proto-neutron star at the bounce shock and the resulting gravitational birth mass of its output compact remnant neutron star. A plot of birth gravitational mass versus compactness from \citet{Burrows2024}
is given in Figure \ref{fig:polyfit}.

\begin{figure} [htbp]
    \centering
    \includegraphics[width=\linewidth]{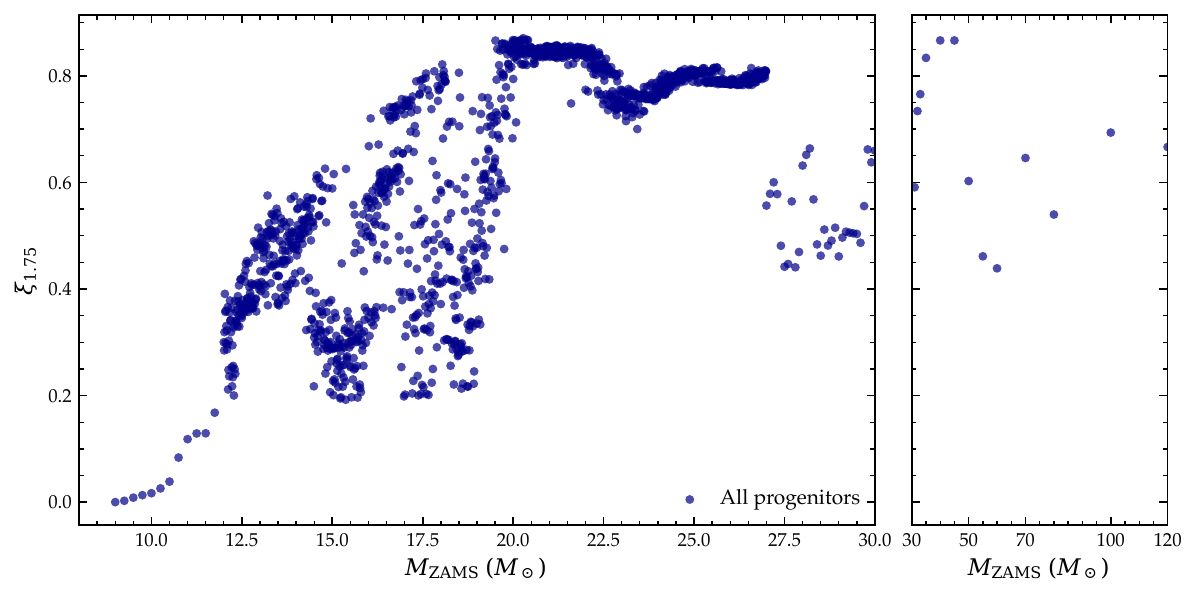}
    \caption{Distribution of core compactness at 1.75 M$_{\odot}$ versus progenitor ZAMS mass, from 9-120 M$_{\odot}$, derived from the solar-metallicity massive star model set of \protect\citet{swbj16,sukhbold2018}. This illustrates the phase space of models and parameters that are our baseline.}
    \label{fig:eta}
\end{figure}

The mapping between compactness and ZAMS mass is still a subject very much in flux, but is certainly quite non-monotonic.  An example from the stellar evolution calculations of \citet{swbj16,sukhbold2018} is shown in Figure \ref{fig:eta}.

We note that the progenitor-to-remnant relation alone requires an IMF to provide an estimate of the number of stars for any given ZAMs mass range. In our baseline calculations, we therefore weight the progenitor side using an initial mass function: for instance, the Salpeter IMF, 
\begin{equation}
\phi(M) \propto M^{-2.35}\, ,
\label{salpeter}
\end{equation}
although our algorithm also permits alternative mappings over specified ZAMs mass ranges \citep{salpeter,woosley2020,Meskhi2022}. 

In implementation, the discrete ZAMs grid is first converted into adjacent, non-overlapping intervals
\[
I_i = 
[M_{i, \mathrm{left}}, \,M_{i, \mathrm{right}})\, ,
\]
whose boundaries are defined by the midpoints between neighboring ZAMs grid points and then clipped to the active progenitor domain. This midpoint construction is numerically significant, as it ensures that the discrete stellar models partition the ZAMS axis without overlap, preventing densely sampled regions of the progenitor grid from being double counted and preserving the multivalued \(M_{\rm ZAMS}\rightarrow\xi_{1.75}\) structure.

Each ZAMS interval is assigned an IMF weight by integrating the chosen IMF across that full interval,
\vspace{0.1cm}

\begin{equation}
\widetilde{w}_i
=
\int_{M_{i, \mathrm{left}}}^{M_{i, \mathrm{right}}} \phi(M)\,dM\, ,
\label{weight}
\end{equation}
and these interval weights are then normalized over the full active progenitor range \([M_{\min},M_{\max}]\),

\begin{equation}
w_i 
= \frac{\widetilde{w}_i}
{\displaystyle
\int_{M_{\min}}^{M_{\max}}\phi(M)\,dM},
\qquad
\sum_i w_i = 1\, ,
\label{normal}
\end{equation}
after masking the mapping to account for potential black hole formation, after which the weights are renormalized appropriately.  Thus, each finite-width progenitor interval contributes exactly one weighted point in the gravitational birth mass space. The model notably does not spread the weight of interval \(I_i\) continuously over a range of remnant masses, and instead deposits the entire normalized interval weight \(w_i\) at the single mapped birth mass \(m_i\). In this sense, the theoretical distribution is a weighted sum  of point-mass contributions:

\[
\mathcal{P}_{\mathrm{th}}(m) 
= 
\sum_i w_i\, \delta(m - m_i)\, .
\]
For visualization, these weighted point contributions are accumulated into display-only birth mass histogram bins \(L_k\) according to
\[
H_k = 
\sum_i w_i\, \mathbf{1}_{L_k}(m_i)\, ,
\]
where \(\mathbf{1}_{L_k}(m_i)=1\) if \(m_i\in L_k\), \(0\) otherwise (an indicator function). The displayed histogram does not exactly represent the theoretical object itself, but instead provides a coarse rendering of the underlying weighted delta-function-like set of contributions. This weighted point structure enables the mapping between the IMF's continuous weighting over each interval and the discrete remnant picture in which there is only one mapped birth mass point per interval. 

In summary, we combine a ZAMS mass function (such as Salpeter) with a compactness to ZAMS mass mapping (such as seen in Figure \ref{fig:eta}) and a compactness to gravitational mass function (for example, what is plotted in Figure \ref{fig:polyfit}) to derive a theoretical birth mass function for neutron stars. This can be compared with measured and inferred neutron star mass functions (\S\ref{measured_1}) to constrain theory. Something similar was performed by \citet{raithel}, but they did not use a relationship between compactness and gravitational mass derived from 3D supernova simulations. Instead, they used the \citet{swbj16} prescription, which relied on artificial 1D explosions, anchored by ansatz using explosion energies from those empirically inferred for the Crab and SN1987A. Moreover, the \citet{swbj16} were unable to birth neutron stars with masses above 1.7 $M_{\odot}$. One of the first theoretical investigations into the neutron star birth mass function was performed by \citet{Timmes1996}. They identified two very narrow peaks that they associated with progenitors which either ignited core carbon burning radiatively or convectively. However, their conception of CCSN explosions was intuitive and not based on detailed supernova modeling.

\subsection{Removing Black-Hole-Forming Candidates}
\label{BH_removal}

\begin{figure} [htbp]
    \centering
    \includegraphics[width=\linewidth]{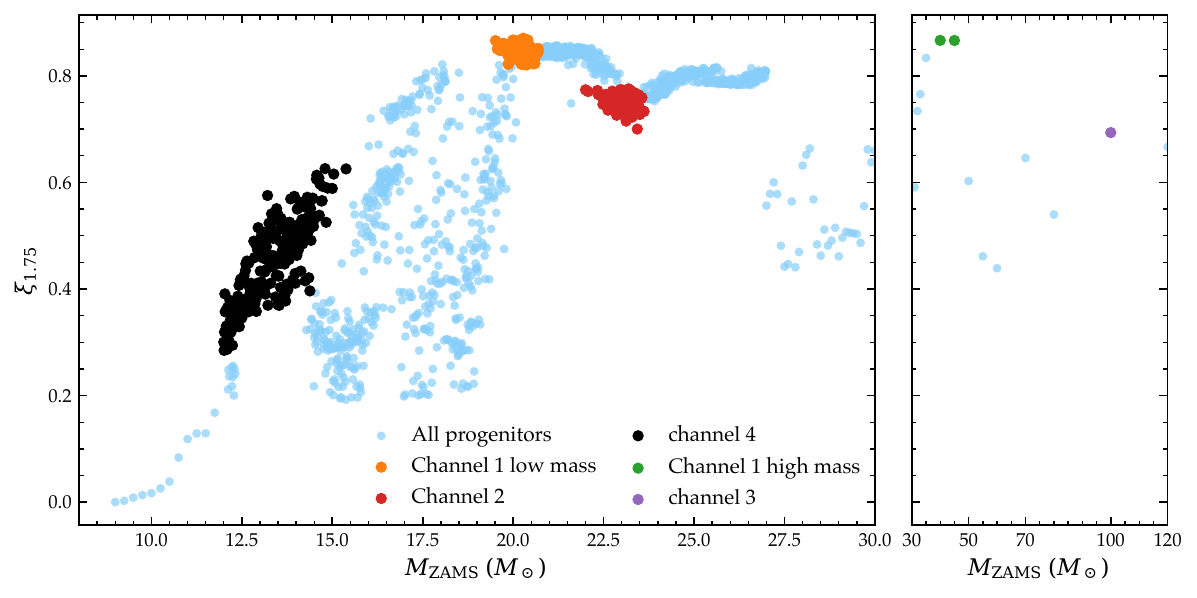}
    \caption{Approximate locations of the four channels of black hole forming regions in the $M_{\rm ZAMS}$ vs. $\xi_{1.75}$ plane according to~\citet{Burrows2025}. The low-mass component of channel 1 is in orange, and the high-mass component is in green. Channel 2 is in red, channel 3 in violet, and channel 4 is in black. See text for a discussion.}
    \label{fig:channels}
\end{figure}

In order to incorporate the effect of black hole formation on the neutron star birth function, we mask a fraction of the progenitors in the $M_{\text{ZAMS}} -\xi_{1.75}$ plane for various progenitor intervals. These progenitor intervals, also known as black hole forming channels, were taken from \cite{Burrows2025}. Their locations in $M_{\text{ZAMS}}- \xi_{1.75}$ plane are shown in Figure~\ref{fig:channels}, with the low-mass component of channel 1 in orange, the high-mass component in green, channel 2 in red, channel 3 in violet, and channel 4 in black.

Instead of adopting a fixed black-hole-removal fraction by hand, we treat the 
removal fractions in the relevant channels as free parameters, except for channel 3. Channel 3 has only one black hole forming candidate, and we assume it always forms a black hole. 

To incorporate the possibility of black-hole formation into the calculation of the neutron-star birth-mass function, we assign each black-hole-forming channel identified in Figure~\ref{fig:channels} a fractional removal parameter. This parameter represents the fraction of progenitors within that channel that form black holes and therefore do not contribute to the neutron-star birth-mass distribution. The removal fractions associated with the individual channels are treated as free parameters and constrained through a likelihood-based optimization, in which we determine the combination of fractions that provides the best agreement between the predicted birth-mass function and the accretion-corrected observed neutron-star mass distribution. The details of this procedure are described below.

Let $f_{\mathrm{BH},c}$ denote the black-hole removal fraction associated with channel (c). The complete set of channel-dependent removal fractions may then be expressed as the parameter vector
\begin{equation}
\mathbf{f}_{\rm BH}
=
\{f_{{\rm BH},c}\},
\qquad
0 \leq f_{{\rm BH},c} \leq 1 .
\end{equation}

Let $D_c$ denote the set of progenitor intervals whose 
representative ZAMS models are labeled as black-hole-forming in channel $c$. For a given trial vector $\mathbf{f}_{\rm BH}$, the code applies the channel 
masks sequentially. In channel $c$, a subset $S_c \subseteq D_c$ is selected 
such that

\begin{equation}
|S_c| \simeq f_{{\rm BH},c}\, |D_c| ,
\end{equation}
excluding objects that have already been removed by an earlier channel. In the birth function calculation, the weights for the progenitors that belongs to set c are suppressed as,
\begin{equation}
w_i' =
\begin{cases}
0, & i \in \bigcup_c S_c,\\
w_i, & i \notin \bigcup_c S_c .
\end{cases}
\end{equation}
The surviving weighted ZAMS intervals are then mapped through the $\xi_{1.75} \rightarrow M_{\text{NS}}$ relation to construct the corresponding theoretical 
neutron star birth function. After this suppression step, the weights are re-normalized so the new theoretical birth function satisfies
\begin{equation}
    \sum_i w_i^{'} = 1
\end{equation}
over the surviving set.

To determine the preferred channel-removal fractions, we compare each trial  model with the observed neutron-star mass distribution using a likelihood-based  technique. The theoretical birth function is re-binned into the same mass bins as  the observed sample. If $n_k$ is the observed number of neutron stars in mass  bin $k$, and $p_k(\mathbf{f}_{\rm BH})$ is the corresponding model probability  in that bin, we minimize the multinomial negative log-likelihood as,

\begin{equation}
-\ln \mathcal{L}(\mathbf{f}_{\rm BH})
=
-\sum_k n_k \ln p_k(\mathbf{f}_{\rm BH}) .
\end{equation}

Equivalently, we report the likelihood per observed object, 
\begin{equation}
    \rm NLL_{\rm obj} = -\frac{1}{N_{\rm obs}} \sum_k n_k \ln p_k (\mathbf{f}_{\rm BH}),
\end{equation}
where, 
\begin{equation}
    N_{\rm obs} = \sum_k n_k
\end{equation}
is the total number of observed objects.
The best-fitting black-hole-removal prescription is the vector 
$\mathbf{f}_{\rm BH}$ that minimizes this quantity. In this way, the relative  importance of the different black-hole-forming channels is inferred from the  agreement between the predicted birth-mass function and the observed neutron-star mass distribution, rather than imposed a priori.

Note that this prescription says nothing about whether black hole formation is accompanied by a supernova \citet{Burrows2025}. Note also that this approach limits the results to a statistical study in remnant survival. In any case, one can apply to the compactness/ZAMS-mass relation (Figure \ref{fig:eta}) any mask representing black hole formation when deriving the neutron star birth mass function. We explore the effect of such mappings in \S\ref{results}, informed in part by (but not limited to) what has been suggested in \citet{Burrows2025} concerning the different potential channels of black hole formation in the context of core collapse. 

\subsection{Post-Formation Accretion}
\label{accretion}

\citet{Zhang2011} proposed a relationship between spin period and accreted mass, showing that shorter spin periods correspond to larger accreted masses (see also \citet{hu2026}).
They derived the following approximate relation, 
\begin{equation}
\Delta M \approx 0.43\,M_\odot \left(\frac{P}{1\,\mathrm{ms}}\right)^{-2/3}\, .
\label{mspin}
\end{equation}
Evaluated across the recycled-pulsar regime, this relation implies accretion of order a few 0.1 $M_{\odot}$ near \(P\sim1\) ms, of order 0.1 $M_{\odot}$ by \(P\sim10\text{--}20\) ms, and well below that scale by \(P\sim100\) ms. Due to this, they suggested that pulsars with spin periods below \(20\) ms have a mean mass of \(1.57\pm0.35\,M_\odot\) whereas slower pulsars have a mean mass of \(1.37\pm0.23\,M_\odot\). 


In 2016, \citet{antoniadis}, modeled millisecond pulsars with a Bayesian framework and used Markov chain Monte Carlo to compare a single Gaussian, a bimodal Gaussian mixture, and an accretion-informed model. \citet{antoniadis} suggested that the single-Gaussian model does not describe the millisecond pulsar (MSP) population adequately, whereas the preferred bimodal Gaussian mixture contains a low-mass component centered at \(1.393^{+0.031}_{-0.029}\,M_\odot\) with width \(0.064^{+0.064}_{-0.025}\,M_\odot\) and a high-mass component centered at \(1.807^{+0.081}_{-0.132}\,M_\odot\) with width \(0.177^{+0.115}_{-0.072}\,M_\odot\). They also derive a lower limit of \(M_{\max}\ge 2.018\,M_\odot\) at \(98\%\) confidence from the absence of a high-mass truncation in the observed MSP masses. This accretion-informed model was used to determine whether the heavy MSP population could be explained by recycling from an underlying birth distribution, whilst their bimodal Gaussian mixture allows for the formation of a distinct heavy component in their intrinsic mass distribution. \citet{antoniadis} concluded that accretion onto MSPs is insufficient to account for the observed large-mass range of neutron stars, and that the heavier MSP systems are therefore unlikely to be produced by recycling alone. This suggested that there remains a substantial range in birth mass, even after accretion is considered.

More recent work by \citet{You2025} addressed the problem of the mapping of period to accreted mass by evolving recycled pulsar systems backward in the \(P\)–\(\dot P\) plane and then applying subclass-dependent accreted-mass prescriptions. They modeled the accreted masses with two prescriptions depending on the recycled subclass. For one group, they applied  a simple scaling relation \(\Delta m/M_\odot = 0.2\,P_{\rm ms}^{-1}\), where \(P_{\rm ms}\) is the spin period in milliseconds, while for the other they used a relation with mean \(\mu_M/M_\odot = 0.3\,P_{\rm ms}^{-1/2}\) and a \(20\%\) dispersion, motivated by numerical simulations of the recycling process \citep{You2025}. They also noted that while recycling may increase neutron-star mass by up to about \(0.3\,M_\odot\), their fitted birth-mass distribution remains concentrated strongly at lower masses once these corrections are applied. \citet{You2025} thus explicitly connected period, spin-down evolution, and accreted mass to arrive at a birth function that is built on a spin-aware correction, rather than on a static mass subtraction. They found that the birth masses are described by a unimodal turn-on power-law distribution that turns on at \(1.10^{+0.04}_{-0.05}\,M_\odot\), peaks at \(1.27^{+0.03}_{-0.04}\,M_\odot\), declines with power-law index \(6.47^{+1.28}_{-1.15}\), and is strongly favored over the widely adopted empirical double-Gaussian model at roughly the \(3\sigma\) level. They also found only marginal evidence for a maximum-mass cutoff, with \(m_{\max}=2.36^{+0.29}_{-0.17}\,M_\odot\), and no evidence for additional peaks on top of the power-law function.

These studies all imply that accretion should be accounted for when comparing predicted a theoretical birth mass function with the observed sample, but suggest that the observed high-mass regime can't be attributed to accretion alone \citep{Zhang2011,antoniadis,You2025}. They also suggest a range of mappings between birth and measured mass which are functions of current period and each provide a crude relationship between mass accreted after birth and the observed spin period. In any case, all these studies suggest that the birth mass function must have a tail to high masses, consistent with the results of \citet{Burrows2024} and Table 1 in the Appendix.

\section{Results}
\label{results}

We now proceed to compare our theoretical predictions for the neutron star mass function from supernova theory with the observed mass function. Our comparison starts with the default birth function, which is first uncorrected by the effects of subsequent mass accretion, the alternative of black hole formation, and the potential effects of kicks on binary survival. We then address in turn the potential character and magnitude of 
black hole masks (\S\ref{black-hole-correction}), followed by an analysis of the alterations due to subsequent mass accretion from a surviving companion (\S\ref{accretion-correction}). We then suggest further modifications to the theoretical mass function due to the potential effect of differential unbinding of a binary suggested in the theoretical neutron star birth mass dependence of neutron star kicks (\S\ref{kick_correction}) \citep{spin2023,Burrows2024}. After these explorations, we provide our resulting overall best-fit theoretical neutron star mass function and compare it with the observations (\S\ref{final}). We emphasize that the primary goal of this best-fit distribution is to show the significance of various correction effects on the birth function for it to resemble the present-day measured mass function of neutron stars. We conclude this section in \S\ref{issues} with a discussion of some outstanding issues.  

\subsection{Default Birth function}
\label{default}

In this section, we construct the default theoretical neutron-star birth-mass function using the mapping $M_{\rm ZAMS} \longrightarrow \xi_{1.75} \longrightarrow M_{\rm birth}$, where $M_{\rm birth}$ denotes the neutron-star mass at formation. The methodology used to establish this mapping is described in detail in \S\ref{method}. The relation between the zero-age main-sequence mass, $M_{\rm ZAMS}$, and the progenitor compactness, $\xi_{1.75}$, is obtained from the stellar-evolution models of \citet{swbj16} and \citet{sukhbold2018}. The subsequent mapping from $\xi_{1.75}$ to $M_{\rm birth}$ is derived from the suite of three-dimensional core-collapse supernova simulations presented by \citet{Burrows2024}.

Various collaborations have recently evolved in three-dimensions the convective silicon- and oxygen-burning regions of a massive star during the final minutes before core collapse \citep{2013ApJ...778L...7C,2015ApJ...808L..21c,muller2016,2020ApJ...890...94Y,muller_lowmass,takahashi2019,fields2020,2021arXiv210704617F,2021ApJ...908...44Y,2021MNRAS.506L..20Y,2021arXiv210700173M,2020MNRAS.497.4644M,varma21}. Their goal is to hydrodynamically model the asymmetries and turbulence of the progenitor pre-collapse and to determine the effects of 3D progenitor charactertistics on the supernova explosion itself \citep{muller2017,Muller2019,bollig2021}. However, detailed multi-dimensional core-collapse simulations of multi-dimensional progenitors across a broad continuum of massive star supernova progenitors are still lacking \citep{vartanyan2022}. As a result, one must fall back on the comprehensive \citet{swbj16} and \citet{sukhbold2018} collection of 1D progenitors to address the systematics of supernova outcomes on progenitor characteristics. This is what we have done here, but the reader should be aware that the issue of the compactness/progenitor-mass mapping is still an open question.

To convert these progenitor-to-remnant relations into a neutron-star birth-mass distribution, each progenitor model is weighted according to a Salpeter initial mass function, $dN/dM_{\rm ZAMS}\propto M_{\rm ZAMS}^{-2.35}$ \citep{salpeter}. The resulting distribution represents our baseline theoretical prediction before accounting for additional effects such as black-hole formation, mass accretion, and natal kicks. We subsequently compare this default birth function with the observed neutron-star mass distribution to identify the principal discrepancies between theory and observation and to determine which physical corrections may be required. 

To compare the theoretical birth-mass distribution with observations, we consider two observational samples: the full set of neutron stars with measured masses and a restricted subset containing only objects with measured spin periods. The latter sample is included because the accretion corrections introduced in subsequent sections depend on the spin period and can therefore be applied only to neutron stars for which this quantity is available. We also exclude the X-ray binaries 4U 1538$-$522 and Her X$-$1 from the subsequent analysis, making the total number of observed neutron star count 146. Their reported masses, $1.02 \pm 0.17 \, M_\odot$ and $1.07\pm 0.36 \, M_\odot$, respectively, are subject to substantial uncertainties. Consequently, their placement in the $1.0-1.2\,M_\odot$ range is not robust evidence that either neutron star has a mass below 1.1 $M_\odot$, but may instead reflect the broad mass posteriors extending into the low-mass regime.

\begin{figure} [htbp]
    \centering
    \includegraphics[width=\linewidth]{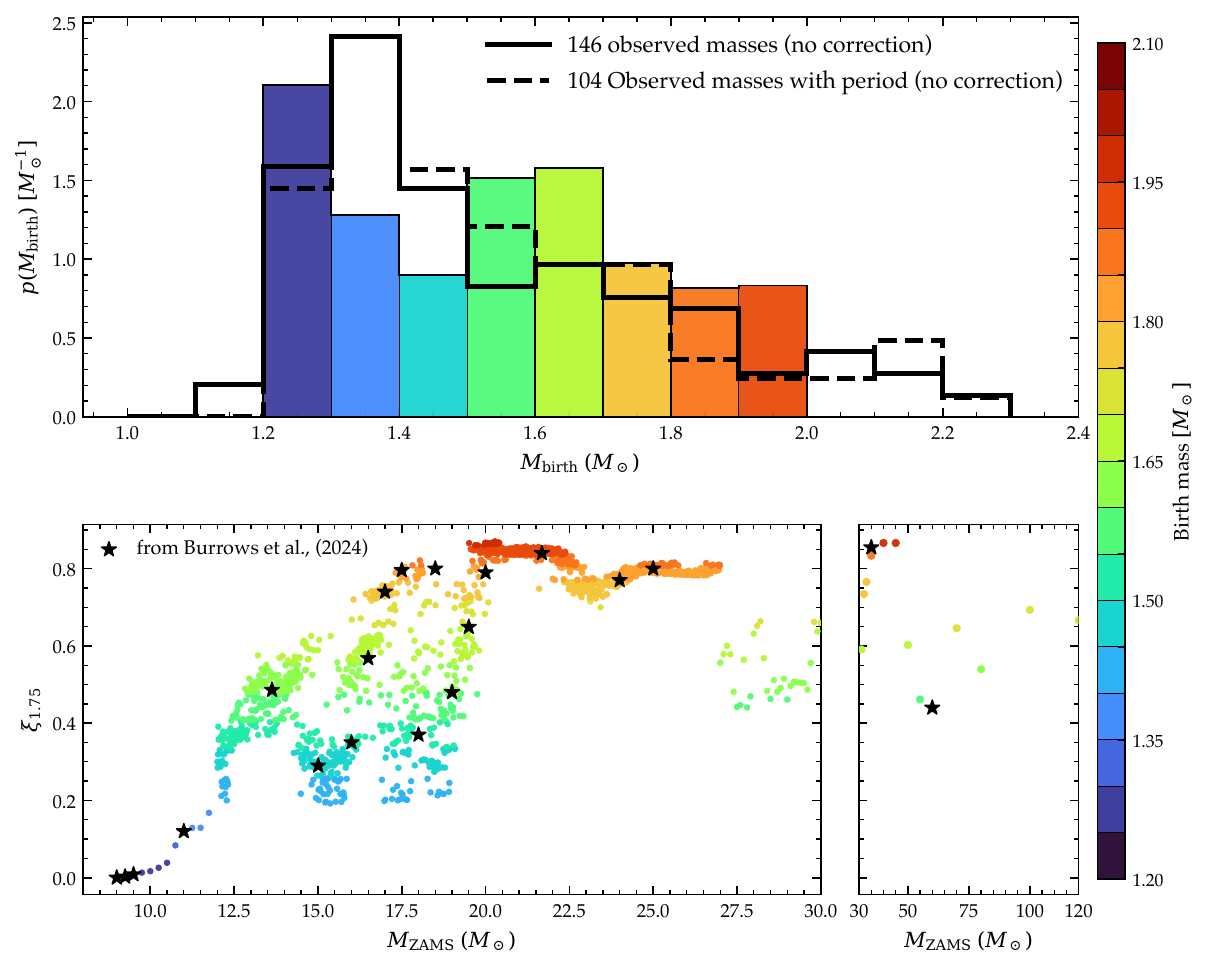}
    \caption{The default theoretical birth mass function for neutron stars calculated using the method in \S\ref{method}. Each bin in this colored histogram is color-coded as a function of birth mass according to the color map on the right. The observed mass distribution of all 146 neutron stars is shown in the solid black line, and the same for 104 neutron stars with period measurements is shown in the dashed black line. All histograms are binned with a 0.1$M_\odot$ bin width. The bottom panel shows the  $M_{\rm ZAMS}$ vs. $\xi_{1.75}$ relation, where each point is color-coded according to the mass bin to which it contributes in the theoretical birth function. The progenitor models used for core-collapse supernova simulation to derive the mapping between compactness and neutron-star birth mass in \citet{Burrows2024} are marked. Noticeably, the observed distributions peak in the range $1.3-1.4\,M_\odot$, whereas the theoretical mass function peaks in the range $1.2-1.3\,M_\odot$. Also, the theoretical mass function shows a sharp minimum and maximum mass cutoff at 1.2 $M_\odot$ and 2.0 $M_\odot$, respectively, which is not reproduced in the observations.}
    \label{fig:birth-function-no-correction}
\end{figure}

Figure~\ref{fig:birth-function-no-correction} presents the default theoretical neutron-star birth mass function, shown in the upper panel. Here, each histogram bin is color-coded according to neutron-star birth mass, as indicated by the accompanying color bar. For comparison, the observed neutron-star mass distribution, without any correction for mass accretion, is shown for two samples (black lines). The solid black curve represents all 146 neutron stars with measured masses, as discussed in \S\ref{measured_1}, whereas the dashed black curve represents the subset of 104 neutron stars for which both mass and spin-period measurements are available.

The lower panel shows the relation between $M_{\rm ZAMS}$ and $\xi_{1.75}$, also presented in Figure~\ref{fig:eta}. Each progenitor is color-coded according to the neutron-star birth-mass bin to which it contributes in the upper panel, thereby directly connecting features in the theoretical birth function to specific regions of the progenitor mass-compactness relation. As expected, the progenitors with the lowest compactnesses, located in the $M_{\rm ZAMS}\simeq 9$-$10.5\,M_\odot$ range and shown in deep blue, contribute primarily to the lowest neutron-star birth-mass bin, $1.2-1.3\,M_\odot$. This relatively narrow progenitor-mass interval contains approximately $20\%$ of the total Salpeter-IMF weight between $9$ and $120\,M_\odot$. Its large statistical weight, combined with the low compactnesses of these progenitors, therefore causes the $1.2-1.3\,M_\odot$ bin to form the dominant peak of the theoretical birth-mass distribution.

The theoretical birth function also exhibits a secondary enhancement in the $1.5-1.7\,M_\odot$ range. This feature is produced mainly by progenitors with compactnesses between approximately $0.4$ and $0.7$, shown in light green in the lower panel. A substantial fraction of these progenitors lie in the $M_{\rm ZAMS}\simeq 12-15\,M_\odot$ interval. Notably, this progenitor-mass range overlaps with one of the stellar-mass black-hole formation channels identified by \citet{Burrows2025}. The consequences of removing or suppressing this channel in the theoretical neutron-star birth function are discussed in \S\ref{black-hole-correction}.

The most massive neutron stars in the theoretical distribution, with birth masses in the $1.8-2.0\,M_\odot$ range, are produced by progenitors located near local maxima of the compactness relation. These progenitors occupy several distinct intervals in initial mass, namely $M_{\rm ZAMS}\sim 19-22.5\,M_\odot$, $24.5-27.5\,M_\odot$, and $35-40\,M_\odot$. Thus, the highest-mass neutron stars are not formed exclusively by the progenitors with the largest values of $M_{\rm ZAMS}$. Instead, they arise from several separated progenitor-mass intervals associated with peaks in $\xi_{1.75}$. This behavior further illustrates the strongly non-monotonic nature of the $M_{\rm ZAMS}$-$\xi_{1.75}$ relation.

The observed mass distribution of the full sample of 146 neutron stars shows only a small variation from the observed mass distribution of 104 neutron stars with spin periods. The most noticeable difference is a small pile-up of neutron stars in the $1.1-1.2 \, M_\odot$ bin for the full 146 sample, which is absent from the 104 neutron star sample boasting spin periods. Moreover, in the $1.5-1.6\, M_\odot$ bin the spin-period-measured sample overestimates the probability density by $\sim 25\%$ compared to the full observed sample. Another noticeable difference occurs in the $1.8-1.9\, M_\odot$ bin, where the spin-period-measured sample underestimates the probability density by $\sim 35\%$ compared to the full observed sample. Despite these differences, both distributions follow the same overall trend of having a peak in the $1.3-1.4\, M_\odot$ bin and then gradually fading away as they approach higher neutron star mass. Therefore, in the following discussion of this subsection, we compare the default theoretical birth function with the spin-period-measured sample of 104 neutron stars. However, a very similar comparison would hold between the default theoretical birth function and the full observed sample of 146 neutron stars. Also, in the discussion that follows, for simplicity we refer to the spin-period-measured sample as the ``observed sample."

The most prominent difference between theoretical birth function and the observed mass distribution appears in the location of the dominant peak. In the observed distribution, the primary peak lies in the $1.3-1.4\,M_\odot$ bin, whereas the theoretical birth function peaks at lower masses, in the $1.2-1.3\,M_\odot$ bin. This offset is naturally expected if a substantial fraction of neutron stars gain mass through accretion after birth. In that case, the present-day observed masses would be systematically shifted upward relative to the underlying birth-mass distribution. The role of accretion in reconciling the locations of these two peaks is examined in detail in \S\ref{accretion-correction}.

The observed mass distribution flattens out over the $1.6 - 1.8 \, M_\odot$ range, whereas the theoretical birth function exhibits a secondary peak near this region. This secondary peak is a feature of the theoretical birth function alone and is not produced in the observed distribution, even after accretion correction (see \S\ref{accretion-correction}). In \S\ref{black-hole-correction} and \S\ref{kick_correction}, we discuss potential correction effects that could remove this discrepancy.

At higher masses, the theoretical birth function predicts a larger number of neutron stars in the $1.8-2.0\,M_\odot$ range than is present in the observed distribution. This excess may indicate that not all progenitors mapped to these high neutron-star birth masses ultimately leave neutron-star remnants. Instead, a fraction of them may undergo black hole formation, either explosively or relatively quiescently \citep{Burrows2025,vartanyan2026_Ni}. The effect of removing such black-hole-forming progenitors from the theoretical neutron-star birth function is explored in \S\ref{black-hole-correction}. In contrast to these overestimates, the theoretical birth function has a sharp cutoff at $M_{\mathrm{birth}} = 2.0\,  M_\odot$, whereas the observed distribution extends up to $2.3 M_\odot$. We discuss this in more detail in \S\ref{accretion-correction} and \S\ref{kick_correction}.

\subsection{Accretion Correction of Observed Neutron Star Mass}
\label{accretion-correction}
In this section, we correct for mass accretion effects on the neutron stars for the observed mass distribution using the prescriptions of \citet{Zhang2011} and \citet{You2025} as discussed in \S\ref{accretion}, and then compare it with the theoretical birth function after accounting for black-hole correction.

\begin{figure} [htbp]
    \centering
    \includegraphics[width=\linewidth]{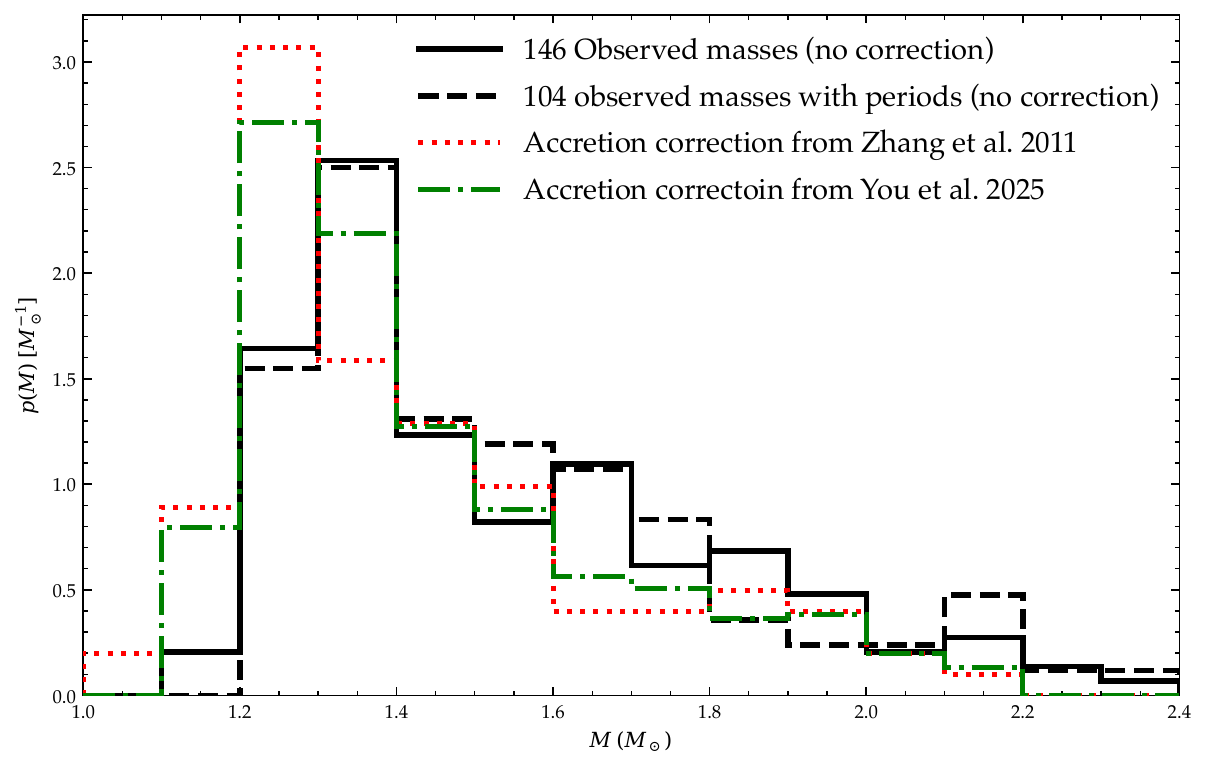}
    \caption{Effect of accretion correction on the observed neutron-star mass distribution using the prescriptions of \citet{Zhang2011} and \citet{You2025}. The solid black curve shows the uncorrected mass distribution of all 146 observed neutron stars, while the dashed black curve shows the uncorrected distribution of the 104 systems with measured spin periods. The accretion-corrected distributions obtained using \citet{Zhang2011} and \citet{You2025} are shown by the red dotted and green dash-dotted curves, respectively. All distributions are constructed using a bin width of $0.1\,M_\odot$. Both correction prescriptions shift the dominant peak from the $1.3-1.4\,M_\odot$ bin to the $1.2-1.3\,M_\odot$ bin. Relative to the uncorrected period-selected distribution, the correction also reduces the probabilities in the $1.6-1.7\,M_\odot$ and $1.7-1.8\,M_\odot$ bins by $8.75\%$ and $3.75\%$, respectively. Finally, the correction slightly contracts the high-mass tail, reducing its upper extent from approximately $2.4\,M_\odot$ to $2.2\,M_\odot$.
}
    \label{fig:accretion}
\end{figure}

Figure~\ref{fig:accretion} illustrates how correcting for post-birth mass accretion may modify the observed neutron-star mass distribution to determine the birth mass function. We apply the accretion-correction prescriptions of \citet{Zhang2011} and \citet{You2025}. The solid black curve shows the mass distribution of the full sample of 146 neutron stars with measured masses, while the dashed black curve shows the distribution of the subset of 104 neutron stars for which spin-period measurements are also available. The accretion-corrected distributions obtained using the prescriptions of \citet{Zhang2011} and \citet{You2025} are shown by the red dotted and green dash-dotted curves, respectively. Because both correction methods depend on the neutron-star spin period, the accretion corrections can be applied only to the period-selected sample of 104 neutron stars. The appropriate baseline for assessing the effect of either correction is therefore the dashed black distribution.

Both accretion-correction prescriptions produce the same principal qualitative effect: probability is redistributed from the observed masses toward lower inferred birth masses. This behavior is expected because the measured mass of an accreting neutron star includes material accumulated after its formation. Removing the estimated accreted mass therefore shifts individual systems, and consequently the overall distribution, toward lower masses. The most visible consequence in both prescriptions is a transfer of neutron stars from the $1.3-1.4\,M_\odot$ bin into the $1.2-1.3\,M_\odot$ bin, causing the dominant peak of the period-selected distribution to move from $1.3-1.4\,M_\odot$ before the correction to $1.2-1.3\,M_\odot$ afterward.

For the \citet{Zhang2011} prescription, the strongest redistribution occurs near the primary peak. Relative to its uncorrected value, the neutron-star count in the $1.3-1.4 \,M_\odot$ bin decreases by a fraction of $0.48$, while the count in the adjacent $1.2-1.3 \,M_\odot$ bin increases by a fraction of $0.92$. This redistribution is sufficiently large to shift the global maximum of the distribution from the $1.3-1.4 \,M_\odot$ bin to the $1.2-1.3 \,M_\odot$ bin. Substantial changes are also present at intermediate masses. The counts in the $1.6-1.7 \,M_\odot$ and $1.7-1.8 \,M_\odot$ bins decrease by fractions of $0.67$ and $0.43$, respectively, relative to their uncorrected values. At higher masses, the largest fractional reductions occur in the $2.2-2.3 \,M_\odot$ and $2.3-2.4 \,M_\odot$ bins. Each of these bins is reduced by a fraction of $1.00$, leaving no neutron stars in either mass interval after the \citet{Zhang2011} correction is applied. At the same time, the $1.8-1.9 \,M_\odot$ bin in the high-mass portion of the distribution increases by a fraction of $0.33$. The cumulative effect of these bin-to-bin transfers is a substantial reduction in the mean neutron-star birth mass. Under the \citet{Zhang2011} prescription, the mean mass of the period-selected sample decreases from $1.55 \,M_\odot$ to $1.43 \,M_\odot$.

The \citet{You2025} prescription produces a qualitatively similar redistribution, although the changes are generally smaller in magnitude than those obtained with the \citet{Zhang2011} prescription. Near the primary peak, the count in the $1.3-1.4 \,M_\odot$ bin decreases by a fraction of $0.12$, while the count in the $1.2-1.3 \,M_\odot$ bin increases by a fraction of $0.75$. As in the \citet{Zhang2011} case, this transfer shifts the dominant peak from the $1.3-1.4 \,M_\odot$ bin to the $1.2-1.3 \,M_\odot$ bin. The intermediate-mass portion of the distribution is also suppressed under the \citet{You2025} correction. The counts in the $1.6-1.7 \,M_\odot$ and $1.7-1.8 \,M_\odot$ bins decrease by fractions of $0.48$ and $0.39$, respectively. These reductions are smaller than the corresponding fractional decreases of $0.67$ and $0.43$ obtained using the \citet{Zhang2011} prescription, but they produce the same overall weakening of the intermediate-mass feature. In the high-mass tail between $2.0$ and $2.4 \,M_\odot$, the four consecutive $0.1 \,M_\odot$ bins decrease by fractions of $0.16$, $0.72$, $1.00$, and $1.00$, respectively. Thus, the suppression generally becomes stronger toward the upper end of the observed mass distribution, and the $2.3-2.4 \,M_\odot$ bin is completely emptied after the correction. Meanwhile, the counts in the $1.8-1.9 \,M_\odot$ and $1.9-2.0 \,M_\odot$ bins increase by fractions of approximately $0.02$ and $0.61$, respectively.

The \citet{You2025} correction consequently reduces the mean neutron-star mass from approximately $1.55 \,M_\odot$ to $1.44 \,M_\odot$, corresponding to a mean downward shift of approximately $0.108 \,M_\odot$. The magnitude of this shift is slightly smaller than the $0.122 \,M_\odot$ shift obtained with the \citet{Zhang2011} prescription, consistent with the generally less pronounced bin-by-bin redistribution seen in Figure~\ref{fig:accretion}.

Although the two prescriptions differ quantitatively, they lead to the same main conclusions. Both shift the dominant peak from $1.3-1.4 \,M_\odot$ to $1.2-1.3 \,M_\odot$, reduce the prominence of the distribution between $1.6$ and $1.8 \,M_\odot$, strongly suppress the highest-mass tail, and lower the mean mass of the period-selected sample by approximately $0.1 \,M_\odot$. For simplicity and consistency, we therefore adopt the \citet{You2025} prescription in the remainder of this work when comparing the theoretical neutron-star birth function with the accretion-corrected observations. The conclusions drawn from these comparisons are expected to remain qualitatively unchanged if the \citet{Zhang2011} prescription is used instead. However, the precise bin-by-bin changes and the inferred magnitude of the correction would differ.

As we have discussed in \S\ref{default}, the full sample of 146 neutron stars and the sample of 104 neutron stars with spin periods have a similar mass distribution. Therefore, in all our subsequent analyses, we use only the spin-period-measured sample for comparison with the theoretical birth function after various correction effects.

\subsection{Black Hole Correction}
\label{black-hole-correction}
We employ the approximate and provisional mapping of ZAMS mass to black hole formation found in the detailed 3D simulation study of \citet{Burrows2025} described in \S\ref{BH_removal}. That paper witnessed the formation of black holes both explosively and quasi-quiescently in four channels.  However, it must be noted that this topic has not been retired; which progenitors form black holes is still very much an open question. Importantly, this study addressed only solar-metallicity models and the dependence upon metallicity has yet to be adequately addressed with modern theoretical tools. Nevertheless, this perspective is usefully informed and viable until a final understanding is achieved.

\begin{figure} [htbp]
    \centering
    \includegraphics[width=\linewidth]{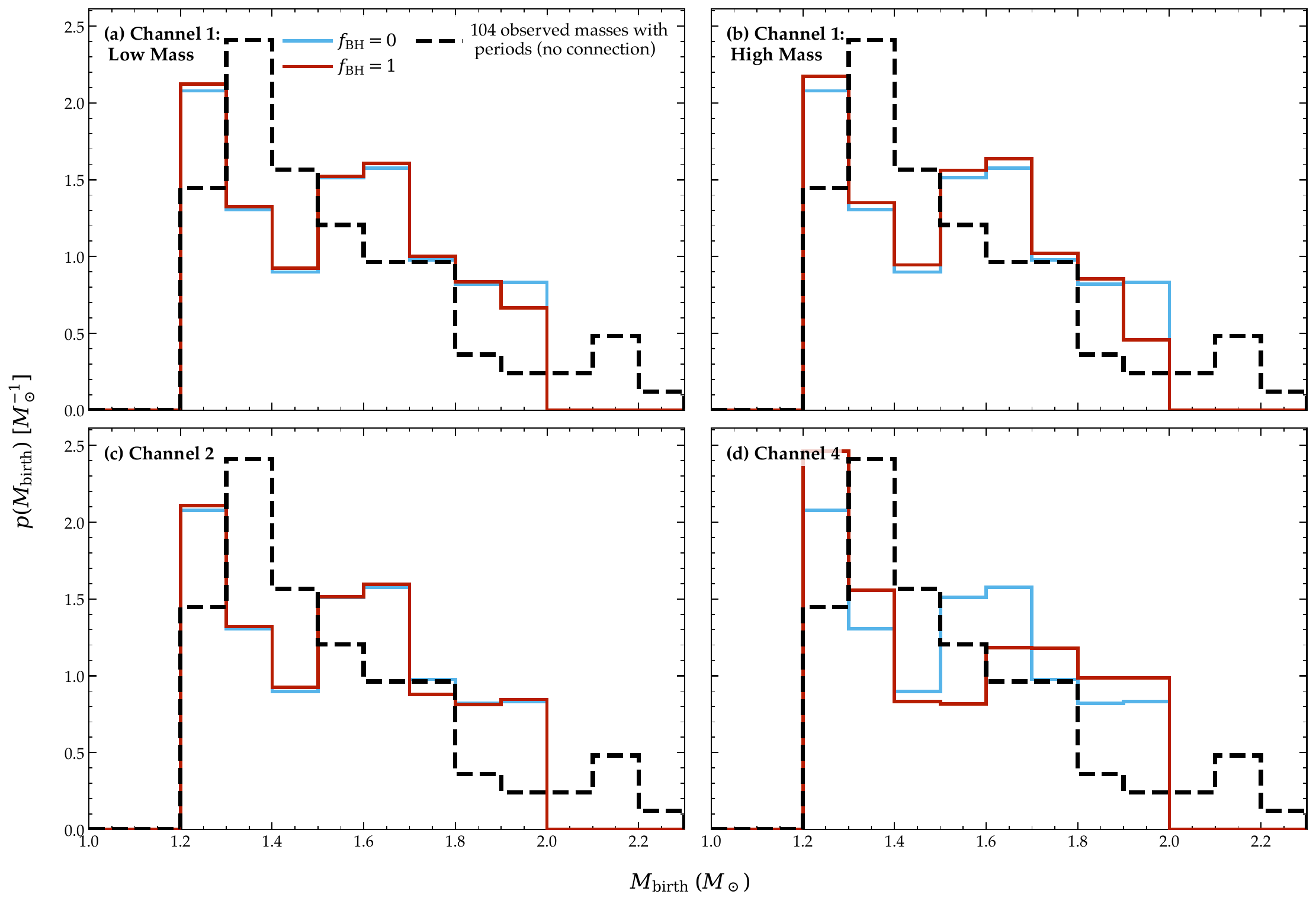}
    \caption{An example of the black-hole masking effect on the theoretical neutron-star birth-mass function. Panels (a)--(d) vary the masking fraction, $f_{\rm BH}$, within the Channel~1 low-mass, Channel~1 high-mass, Channel~2, and Channel~4 black-hole-forming regions identified by \citet{Burrows2025}, respectively. The colored step curves correspond to the values of $f_{\rm BH}$ indicated in the legend. In each panel, the dashed black curve shows the observed mass distribution for the 104 neutron stars with measured spin periods. In panel~(a), increasing the masking fraction from $f_{\rm BH}=0$ to $1$ in the Channel~1 low-mass region reduces the probability in the $1.9-2.0\,M_\odot$ bin by $20.2\%$. In panel~(b), masking the Channel~1 high-mass region produces a larger $45.0\%$ reduction in the same bin. In panel~(c), Channel~2 masking yields a milder $9.9\%$ reduction in the $1.7-1.8\,M_\odot$ bin. Panel~(d) shows the strongest and broadest effect: Channel~4 masking reduces the probabilities in the $1.5-1.6\,M_\odot$, $1.6-1.7\,M_\odot$, and $1.4-1.5\,M_\odot$ bins by $46.0\%$, $24.9\%$, and $7.5\%$, respectively.}
    \label{fig:bh-masking-4panel}
\end{figure}

Figure~\ref{fig:bh-masking-4panel} illustrates how masking progenitors that are expected to form black holes modifies the theoretical neutron-star birth-mass function. The black-hole-forming regions are based on the stellar-mass black-hole formation channels identified by \citet{Burrows2025}. Panels (a)-(d) show the effect of varying the fraction of black-hole-forming candidates, $f_{\rm BH}$, from 0 to 1 within, respectively, the low-mass component of channel~1, the high-mass component of channel~1, channel~2, and channel~4. Channel~3 is not included because it contains only a single progenitor, with $M_{\rm ZAMS}=100\, M_\odot$ (and at $Z = 0.1 Z_{\odot}$), which we assume always forms a black hole.

In each panel, the light-blue solid curve shows the default theoretical neutron-star birth function obtained when none of the progenitors in the selected channel are removed, corresponding to $f_{\rm BH}=0$. The red solid curve shows the opposite limiting case, $f_{\rm BH}=1$, in which all progenitors belonging to that channel are assumed to form black holes and are therefore excluded from the neutron-star birth function.  For comparison, the mass distribution of the 104 neutron stars with measured spin periods is shown by the black dashed curve.

Masking the low- and high-mass components of channel~1 affects the same region of the theoretical birth function, although the magnitude of the suppression differs substantially between the two components. In both cases, the principal change occurs in the $1.9-2.0\,M_\odot$ birth-mass bin. Increasing $f_{\rm BH}$ from 0 to 1 in the low-mass component of channel~1 reduces the probability in this bin by a fraction of $0.20$, whereas removing all black-hole candidates in the high-mass component produces a larger reduction by a fraction of $0.45$. Because the birth function is renormalized after the black-hole-forming progenitors are removed, the suppression of the $1.9-2.0\,M_\odot$ bin is accompanied by increases in the probabilities assigned to the remaining birth-mass bins.

The effect of channel~2 is comparatively mild and is concentrated primarily in the $1.7-1.8\, M_\odot$ bin. Increasing $f_{\rm BH}$ from 0 to 1 reduces the probability in this bin by a fraction of $0.10$. Since channel~2 removes only a relatively small fraction of the total neutron-star-forming weight, the corresponding normalization-induced increases in the other birth-mass bins are very small.

Channel~4 produces by far the largest and most broadly distributed modification to the theoretical birth function. Unlike the other channels, which primarily suppress a single birth-mass bin, channel~4 removes progenitors that contribute across several bins between $1.4$ and $1.7\,M_\odot$. Increasing $f_{\rm BH}$ from 0 to 1 reduces the probability in the $1.4-1.5\,M_\odot$ bin by a fraction of $0.08$, in the $1.5$-$1.6\,M_\odot$ bin by a fraction of $0.46$, and in the $1.6-1.7\,M_\odot$ bin by a fraction of $0.25$.

Because the channel~4 removal eliminates a substantial fraction of the original neutron-star-forming probability, renormalization produces correspondingly large increases in the bins that are not directly suppressed. Relative to their values before black-hole masking, the probability increases by a fraction of $0.19$ in the $1.2-1.3\,M_\odot$ bin, by a fraction of $0.19$ in the $1.3-1.4\,M_\odot$ bin, by a fraction of $0.21$ in the $1.7-1.8\,M_\odot$ bin, by a fraction of $0.20$ in the $1.8-1.9\,M_\odot$ bin, and by a fraction of $0.18$ in the $1.9-2.0\,M_\odot$ bin. The increase is largest in the $1.7-1.8\,M_\odot$ bin, although the changes across all unaffected bins are comparable at approximately $0.18$-$0.21$.

All quoted fractional increases and reductions are measured relative to the original probability in the corresponding birth-mass bin at $f_{\rm BH}=0$. 

Overall, channel~1 black-hole masking primarily reduces the number of neutron stars in the highest-mass bin, $1.9-2.0\,M_\odot$, whereas channel~2 predominantly affects the $1.7-1.8\,M_\odot$ bin, and channel~4 modifies the distribution across the $1.4-1.7\,M_\odot$ range. This channel-dependent imprint of black-hole formation on the neutron-star birth function provides a means of constraining the fraction of black-hole-forming progenitors over the $9-120\,M_\odot$ progenitor-mass range by comparing the black-hole-corrected theoretical birth function with the accretion-corrected observed mass distribution, as explored in the following sections.



\subsection{Results Due to Black-Hole Formation and Accretion Corrections}
\label{BH_formation}

Having constructed the accretion-corrected observed mass distribution for the 104 neutron stars with measured spin periods, we next determine the black-hole masking fractions that allow the theoretical birth-mass function derived in \S\ref{default} to best reproduce the observed distribution. We treat the black-hole formation fractions associated with the three candidate channels, channels 1, 2, and 4, as free parameters, with the low- and high-mass components of channel 1 varied independently. The black hole forming fraction for channel 3 has been fixed to 1 because it has only one progenitor candidate. We then apply the likelihood-based optimization procedure described in \S\ref{BH_removal} to identify the combination of channel-dependent removal fractions that yields the best agreement between the theoretical birth-mass function and the accretion-corrected observations.

\begin{figure} [htbp]
    \centering
    \includegraphics[width=\linewidth]{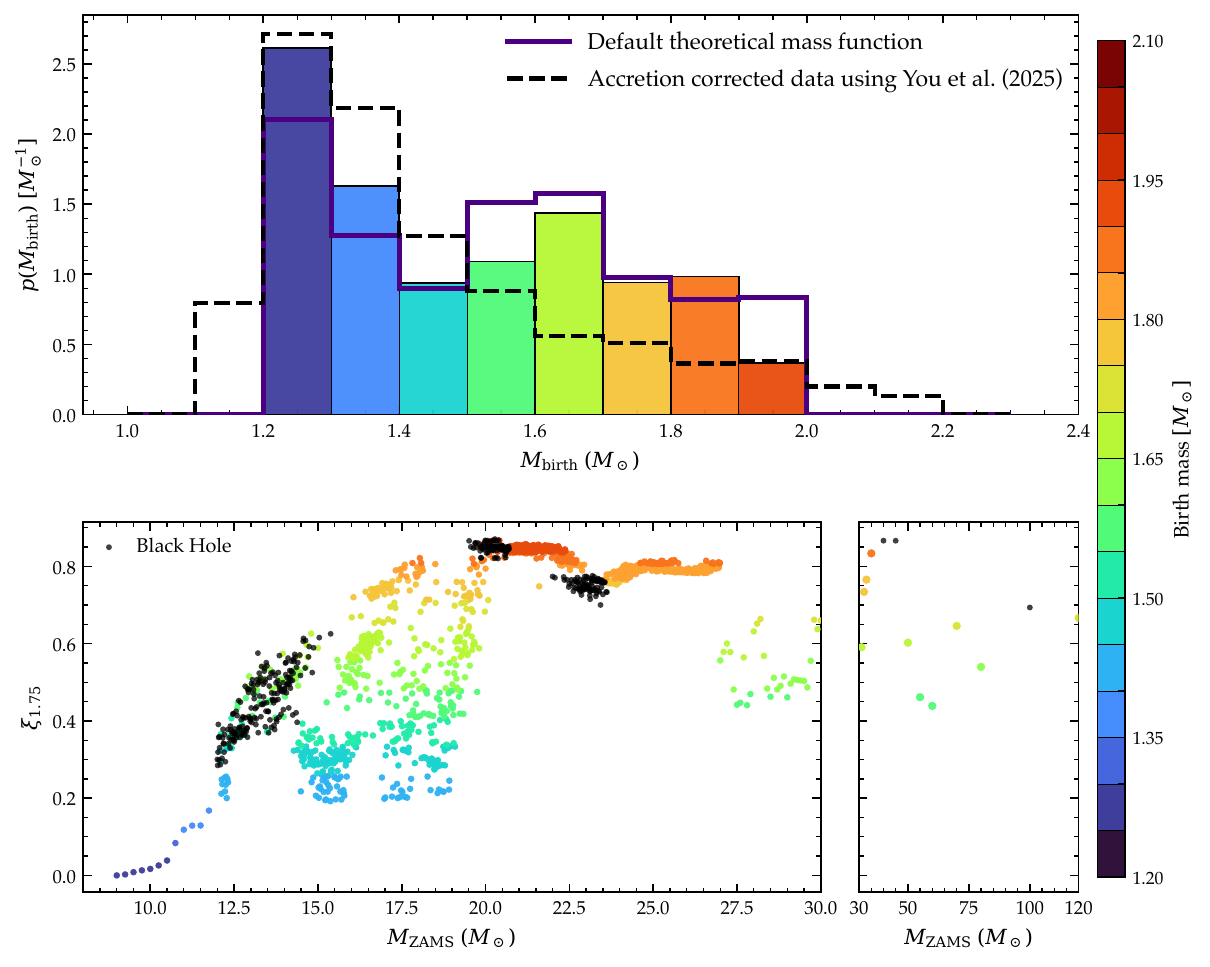}
    \caption{Theoretical neutron-star birth-mass function with best-fitting black-hole masking. The colored histogram in the top panel shows the theoretical birth function after applying the black-hole masking that best fits the observed mass distribution after accretion correction using \citet{You2025} prescription. The bins are color-coded as a function of birth mass as indicated by the color bar. The default theoretical mass function is the purple solid line, and the accretion-corrected observed mass distribution, obtained using the \citet{You2025} prescription, is shown in the dashed black line. The bottom panel shows the $M_{\rm ZAMS}$ vs. $\xi_{1.75}$ relation, with each progenitor colored according to the birth-mass bin to which it contributes. Progenitors removed by the black-hole masking prescription are shown in black. The fraction of black holes removed for each channel is: channel 1 low mass: 0.85, channel 1 high mass: 1.0, channel 2: 1.0, channel 3: 1.0, channel 4: 0.75. The final black hole fraction of progenitors for the best-fit black hole masking is $20.7\%$.}
    \label{fig:bh-mask}
\end{figure}

Figure~\ref{fig:bh-mask} shows the theoretical neutron-star birth-mass function obtained after masking progenitors associated with the black hole formation channels from \citet{Burrows2025}. The masking fractions were determined using the maximum-likelihood procedure described at the end of \S\ref{BH_removal} to obtain the closest agreement with the observed neutron-star mass distribution after applying the accretion correction of \citet{You2025}. In the upper panel, the histogram represents the resulting black-hole-masked theoretical birth function, with each bin color-coded according to birth mass, as indicated by the color bar. For comparison, the default theoretical birth function derived in \S\ref{default} is shown by the solid deep-purple curve, while the accretion-corrected observed mass distribution is shown by the dashed black curve.

The lower panel displays the relation between $M_{\rm ZAMS}$ and $\xi_{1.75}$ for the progenitor models. Each progenitor is assigned the color of the neutron-star birth-mass bin to which it contributes. Progenitors classified as potential black-hole-forming candidates within the approximate channel boundaries defined in \S\ref{BH_removal} are indicated by black points. The adopted masking fractions are $0.90$ and $0.85$ for the low- and high-mass branches of channel~1, respectively, $1.00$ for channels~2 and~3, and $0.75$ for channel~4. After weighting by the initial mass function, these choices correspond to approximately $20.7\%$ of all progenitors in the $9-120\,M_\odot$ ZAMS-mass interval forming black holes rather than neutron stars.

We first compare the black-hole-masked birth function with the default theoretical distribution. The most significant changes occur above approximately $1.5\,M_\odot$. Relative to the default model, black-hole masking reduces the probability density by a fraction of approximately $0.31$ in the $1.5-1.6\,M_\odot$ bin and by a fraction of $0.06$ in the $1.6-1.7\,M_\odot$ bin. As discussed in \S\ref{black-hole-correction}, these reductions arise primarily from channel~4, within which a fraction of $0.75$ of the candidate progenitors are removed. Finally, masking in channel~1 decreases the probability density in the $1.9-2.0\,M_\odot$ bin by a fraction of approximately $0.56$.

Because the distribution is renormalized after the black-hole-forming progenitors are removed, the reductions at intermediate and high birth masses are accompanied by an increase in the relative probability density between approximately $1.2$ and $1.4\,M_\odot$. We also notice an increase in probability weight by a fraction of $\sim 0.18$ after black-hole masking compared to the default birth function in the $1.8-1.9 M_\odot$ bin. This increase is associated with channel-4 masking as discussed in \S\ref{black-hole-correction}. Black-hole masking alone, therefore, does not reproduce the accretion-corrected observed distribution. Nevertheless, it modifies the default theoretical birth function in the required direction: it suppresses several of the intermediate (in $1.5 - 1.7 M_\odot$ bins) and high-mass (in $1.9-2.0 M_\odot$ bin) features that are overproduced by the uncorrected theoretical model and transfers greater relative weight to the low-mass portion of the distribution.

We next compare the black-hole-masked theoretical birth function directly with the observed mass distribution corrected using the prescription of \citet{You2025}. Between $1.2$ and $1.5\,M_\odot$, the theoretical distribution broadly follows the declining trend seen in the corrected observations. Quantitative differences remain, however: the theoretical probability density is lower than the observed value by fractions of approximately $0.04$, $0.27$, and $0.31$ in the three successive bins spanning this mass range compared to the accretion-corrected data. These low-mass discrepancies may indicate that the present progenitor models or remnant-mass prescription do not fully account for all channels capable of producing low-mass neutron stars. Possible explanations are discussed further in \S\ref{issues}.

The largest disagreement occurs between approximately $1.5$ and $1.9\,M_\odot$. Although masking channel~4 partially suppresses the secondary enhancement predicted in this region, it does not remove it completely. Consequently, the black-hole-corrected theoretical distribution retains a pronounced maximum in the $1.6-1.7\,M_\odot$ bin, whereas the accretion-corrected observed distribution decreases approximately monotonically across the same mass interval. A similar disagreement, yet moderate in magnitude, occurs in the $1.8-1.9 M_\odot$ bin, where the mass function increases after accounting for a black-hole correction, in contrast with the gradually decreasing observed distribution. These results suggest that black-hole formation in those four above-mentioned channels alone is insufficient to explain the absence of the predicted intermediate- and high-mass enhancement in the observations. Additional effects may preferentially reduce the number of observable neutron stars in this mass range. These could include binary disruption or selection effects associated with natal kicks, as examined in \S\ref{kick_correction}, as well as possible modifications associated with the initial spin of the newly formed neutron star, discussed in \S\ref{issues}.

At higher masses, the decreasing trend of the theoretical birth function reappears between $1.9$ and $2.0\,M_\odot$, followed by a sharp cutoff near $2.0\,M_\odot$. The black-hole masking substantially reduces the excess predicted by the default model in this region, particularly in the $1.9-2.0\,M_\odot$ bin. Nevertheless, the theoretical distribution still exceeds the accretion-corrected observed probability density in the $1.7-1.9\,M_\odot$ bin by approximately a factor of $2$. As in the intermediate-mass region, this remaining excess may indicate the need for additional corrections beyond the simplified black-hole masking prescription adopted here.

For the present analysis, we have masked only the approximate locations of the four stellar-mass black-hole formation channels identified by \citet{Burrows2025}. These boundaries should not, however, be interpreted as exhaustive. For example, the channel 1 low-mass component is marked as a small cluster of progenitors located at the compactness maximum occurring at $M_{\mathrm{ZAMS}} = 20 M_\odot$ (Figure~\ref{fig:bh-mask} lower panel, points marked in black). However, this compactness maximum stretches horizontally up to $M_{\mathrm{ZAMS}} =22.5 M_\odot$, primarily consisting of red and dark orange points on the lower panel of Figure~\ref{fig:bh-mask}. Due to the progenitors in this horizontal branch having the same compactness as the channel 1 low-mass component, and having even higher $M_{\mathrm{ZAMS}}$, future efforts in 3D CCSN simulations might find these progenitors forming black holes. In fact, the simple parametric explodability prescription of \citet{ertl-2016} and and 1D parameterized explosion model of \citet{mueller-2016} previously identified additional failed explosions near compactness maxima at $M_{\rm ZAMS}\simeq20.5-22\,M_\odot$. In addition,~\cite{ertl-2016} identified possible black-hole formation in the approximate range $M_{\mathrm{ZAMS}} \approx 25-27 \, M_\odot$, associated with the secondary compactness maximum represented by the light orange points in the same figure. These progenitors, colored in red and orange in the lower panel of Figure~\ref{fig:bh-mask}, primarily contribute to the $1.7-2.0\,  M_\odot$ bins in the theoretical neutron star birth function. Masking an appropriate fraction of these additional progenitors would, therefore, further suppress the theoretical distribution precisely where it continues to exceed the accretion-corrected observations. The present result should, therefore, be interpreted as evidence that black-hole formation moves the theoretical birth function toward the observed distribution, while also demonstrating that black-hole masking by itself is unlikely to account for all of the remaining discrepancies.

Finally, the best-fit black hole masking also provides the black hole fraction in the $9-120 \, M_\odot$ progenitor range. For the given black hole masking fractions in different channels in Figure~\ref{fig:bh-mask}, we found $23\%$ of the progenitors formed black holes. After weighting this fraction by the Salpeter mass function~\citep{salpeter}, we obtained the final black hole fraction of $20.7\%$, implying $79.3\%$ of progenitors form neutron stars.

\subsection{Effects of Natal Kicks}
\label{kick_correction}

It is calculated that high compactness (and, hence, high neutron star birth mass) leads, on average and in a statistical sense, to greater kick speeds \citep{spin2023}. Higher kick speeds will, on average, lead to higher binary disruption probabilities \citep{Kalogera1996}. As a result, we posit that the bins for neutron stars at the higher gravitational masses and in binaries will be relatively depleted due to their greater likelihood of becoming unbound, though by how much is uncertain. 

\begin{figure} [htbp]
    \centering
    \includegraphics[width=\linewidth]{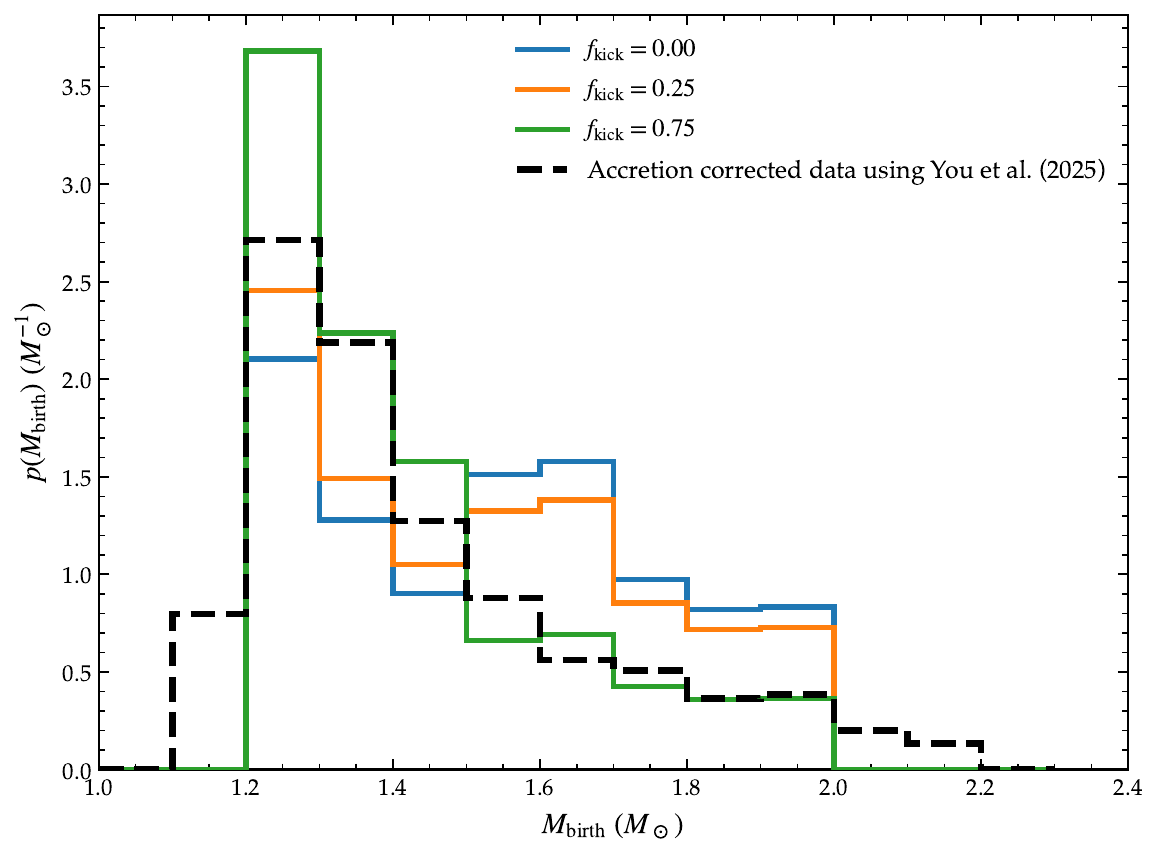}
    \caption{Effect of natal kicks in the mass range between 1.2$M_\odot$ and 2.0$M_\odot$  on the theoretical neutron-star birth-mass function. Each colored curve shows the birth function after removing a specified fraction of remnants from the specified mass range, as indicated in the legend. The black dash-dotted curve shows the observed mass distribution after applying the accretion correction of \citet{You2025}. The comparison indicates that around $75\%$ of the relevant remnants must be removed from the default theoretical birth-mass function to obtain a reasonable fit to the accretion-corrected observed distribution.}
    \label{fig:kick-removal}
\end{figure}

Figure~\ref{fig:kick-removal} illustrates the effect of such natal-kick-induced removal on the theoretical neutron-star birth-mass function. In this simple approach, a fraction $f_{\rm kick}$ of the remnants with birth masses between $1.5$ and $2.0\,M_\odot$ is removed from the theoretical population. We choose this mass range because this is where the disagreement between the theoretical mass function and the observation is the greatest, so we could demonstrate how effects of natal kicks may change the theoretical mass function in the desired way. The reason for this choice will become more clear in the next section, where we statistically calculate the kick fractions in each mass bin. The resulting theoretical birth functions are shown by the solid colored curves, with each color corresponding to a different value of $f_{\rm kick}$. The observed neutron-star mass distribution, corrected for accretion using the prescription of \citet{You2025}, is shown by the dashed black curve. Comparing the sequence of theoretical curves demonstrates how progressively greater removal of intermediate- and high-mass remnants modifies the shape of the birth function and its agreement with the accretion-corrected observations.

The blue curve corresponds to the default theoretical birth function, for which $f_{\rm kick}=0$. As discussed in \S\ref{default}, this distribution exhibits a prominent secondary enhancement between approximately $1.5$ and $1.7\,M_\odot$. No corresponding feature is present in the accretion-corrected observed distribution, which instead decreases with increasing mass throughout this interval. The default theoretical model also overpredicts the probability density in the high-mass tail between approximately $1.8$ and $2.0\,M_\odot$. 

The orange curve shows the result obtained for $f_{\rm kick}=0.25$. Removing $25\%$ of the remnants between $1.5$ and $2.0\,M_\odot$ reduces the probability density in this mass interval by approximately a fraction of $0.15$. Because the surviving distribution is subsequently renormalized, the probability density between approximately $1.2$ and $1.4\,M_\odot$ increases by  a fraction of about $0.14$. This moderate removal fraction partially suppresses both the secondary enhancement and the high-mass excess, thereby shifting the theoretical birth function toward the observed distribution. Nevertheless, the theoretical model continues to substantially overpredict the probability density above $1.5\,M_\odot$, and the secondary intermediate-mass structure remains clearly visible.

As $f_{\rm kick}$ is increased further, the agreement between the theoretical and observed distributions improves progressively. This behavior is particularly evident for $f_{\rm kick}=0.75$, shown by the green curve. Removing $75\%$ of the remnants between $1.5$ and $2.0\,M_\odot$ decreases the theoretical probability density in this interval by approximately $0.61$ in fraction. Renormalization correspondingly increases the probability density between approximately $1.2$ and $1.5\,M_\odot$ by a fraction of about $0.57$. The cumulative effect is a substantial reshaping of the theoretical birth function: the secondary peak between $1.5$ and $1.7\,M_\odot$ is no longer present, the distribution instead develops a single dominant maximum in the $1.2-1.3\,M_\odot$ bin, and the probability density subsequently decreases approximately monotonically with increasing birth mass.

The $f_{\rm kick}=0.75$ model consequently provides substantially better overall agreement with the accretion-corrected observed distribution across the $1.2-2.0\,M_\odot$ mass range. In particular, the suppression of the intermediate-mass enhancement and the reduction of the high-mass tail reproduce the broad shape of the observed distribution more closely than the default theoretical model. Some differences nevertheless remain. The theoretical distribution overpredicts the probability density in the $1.2-1.3\,M_\odot$ bin and exhibits a sharp lower-mass cutoff near $1.2\,M_\odot$, whereas the accretion-corrected observations extend below this value. These remaining low-mass discrepancies are discussed further in \S\ref{issues}.

The treatment adopted here is intentionally notional. Removing different fractions of remnants from a prescribed birth-mass interval does not model the detailed dependence of natal kicks on progenitor structure, explosion dynamics, remnant mass, or binary orbital properties. It therefore cannot replace a self-consistent population-synthesis or dynamical calculation. Nevertheless, this analysis provides a useful estimate of the magnitude of the mass-dependent selection required to reconcile the theoretical birth function with the observed population. It also provides a natural basis for the combined analysis presented in the following section, where we consider the collective influence of black-hole formation and natal-kick removal. We then determine the combination of black-hole-forming and kick-removal fractions that provides the closest overall agreement between the theoretical birth function and the accretion-corrected observed neutron-star mass distribution.

\subsection{Final Aggregate Fit}
\label{final}
Here, we construct the final aggregate fit by accounting for black-hole formation and estimating the neutron-star removal fractions due to kicks needed to bring the theoretical birth-mass function into agreement with observations, and thereby providing approximate targets or detailed binary-unbinding calculations. For simplicity, we mention them as kick fractions, but the reader should keep in mind that they are not a result of a detailed binary disruption calculation. To achieve this, first we modify our likelihood analysis technique mentioned in \S\ref{BH_removal} to also consider a new free parameter $f_{\mathrm{kick}, i} \in [0, 1]$, where $f_{\mathrm{kick}, i}$ stands for the kick fraction in the i-th bin. Then, we jointly estimate the fractions of black-hole-forming progenitors associated with the different formation channels and the kick fractions from each mass bin that provide the closest agreement with the observed mass distribution. Although this approach is simplistic and somewhat notional, our main objective in this study is to understand how one should apply various corrections to the theoretical birth function for it to closely resemble the observed mass distribution. The likelihood analysis technique helps us quantify what part of the theoretical birth function needs to be shaved due to black hole formation and binary unbinding for it to almost reproduce the accretion-corrected observed distribution (using the \citet{You2025} prescription).

\begin{figure}
    \centering
    \includegraphics[width=\linewidth]{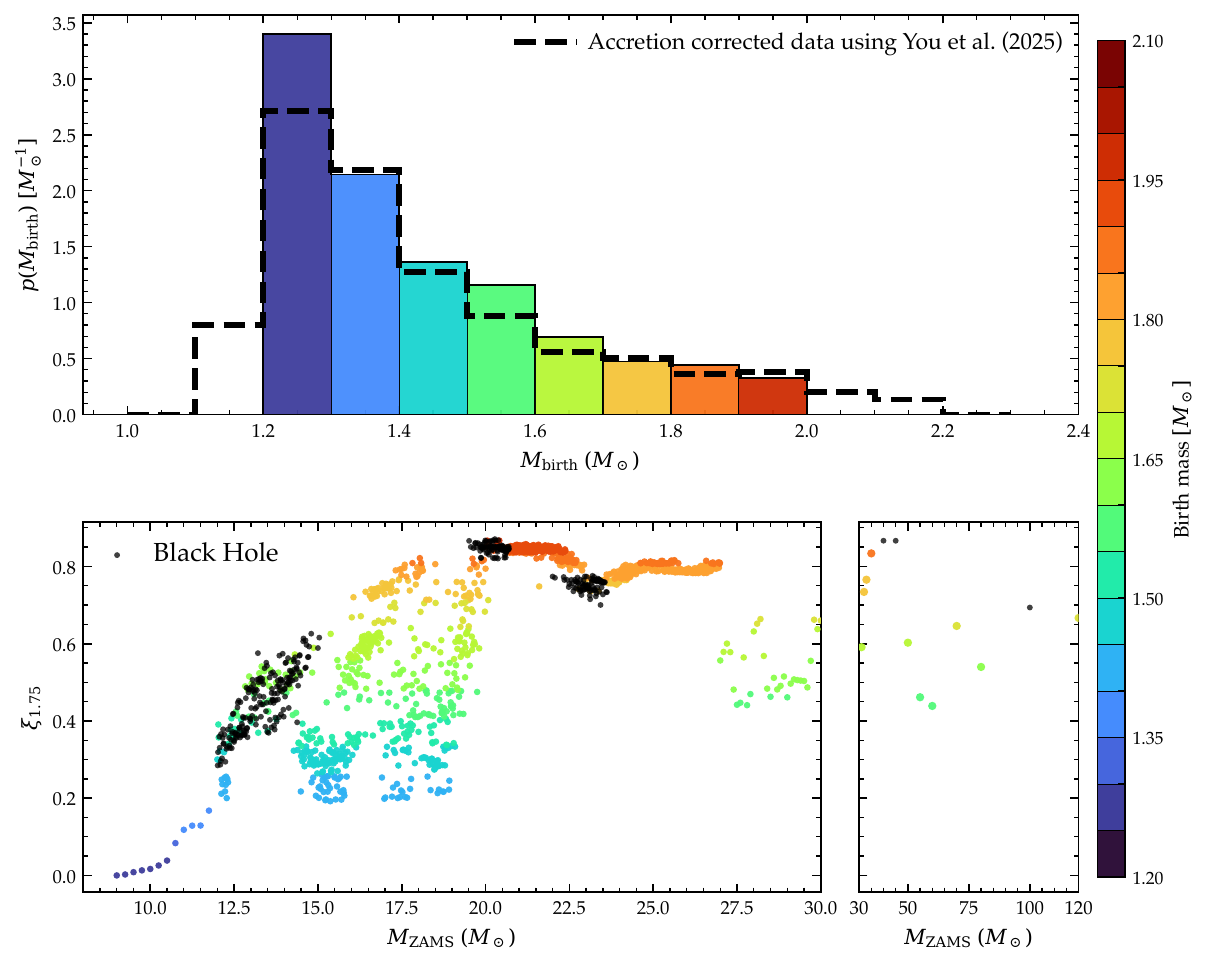}
    \caption{Theoretical neutron-star birth-mass function after applying the best-fit black-hole masking prescription and removing fractions $\{f_{\mathrm{kick}, i}\}$ from the mass bins. For the birth mass bins between $1.2 -2.0 M_\odot$ at $0.1 M_\odot$ intervals, the kick fractions are $\{0.022, 0.023, 0.023, 0.210, 0.631, 0.632, 0.655, 0.312\}$. The fractions of black holes removed for different channels are: channel 1 low mass: 0.9, channel 1 high mass: 1.0, channel 2: 0.9, channel 3: 1.0, channel 4: 0.75. These parameters were selected to best reproduce the accretion-corrected observed mass distribution obtained using the prescription of \citet{You2025}, shown by the dashed black curve. The final fraction of progenitors that form black holes, for the best-fit black-hole masking and kick correction, is $19.6\%$. The colored histogram bins indicate birth mass according to the color bar. The lower panel shows the $M_{\rm ZAMS}$--$\xi_{1.75}$ relation, with each progenitor colored by the birth-mass bin to which it contributes; progenitors removed by black-hole masking are shown in black. The resulting theoretical distribution agrees well with the corrected observations over most of the mass range, but retains a sharp lower cutoff near $1.2\,M_\odot$ that apparently is not present in the data. However, the objects which populate the lowest mass bin have large error bars and we suggest neutron stars are not actually born below 1.1 $M_{\odot}$. It also terminates near $2.0\,M_\odot$, whereas the accretion-corrected observed distribution extends to approximately $2.2\,M_\odot$.}
    \label{fig:you+bh+kick}
\end{figure}

Figure~\ref{fig:you+bh+kick} presents the final theoretical neutron-star birth-mass function with the aggregate fits. In the upper panel, the histogram shows the corrected theoretical birth function, with each bin color-coded according to neutron-star birth mass, as indicated by the color bar. The observed neutron-star mass distribution, corrected for accretion using the prescription of \citet{You2025}, is shown by the dashed black curve. The lower panel displays the relation between $M_{\rm ZAMS}$ and $\xi_{1.75}$ for the progenitor models, with each progenitor assigned the color of the birth-mass bin to which it contributes. Progenitors removed as black-hole-forming candidates within the channels identified by \citet{Burrows2025} are shown as filled black circles.

The adopted black-hole formation fractions are $0.90$ and $1.0$ for the low- and high-mass branches of channel 1, $0.9$ for channel 2, $1.0$ for channel 3, and $0.75$ for channel 4. These masking fractions imply that approximately $19.6\%$ of the progenitors in the $M_{\rm ZAMS}=9-120\,M_\odot$ interval form black holes rather than neutron stars. We then remove a fraction $(f_{{\rm kick},i})$ of neutron stars from each of the i-th birth mass bins. For birth masses spanning $(1.2\leq M_{\rm birth}/M_\odot\leq2.0)$, divided into bins of width $0.1\,M_\odot$, the inferred removal fractions are $f_{{\rm kick},i}=\{0.022, 0.022, 0.023, 0.210, 0.631, 0.632, 0.654, 0.312\}$, listed in order of increasing birth mass. The removal fractions required to closely reproduce the accretion-corrected observed mass distribution tend to increase with neutron-star birth mass. This trend is qualitatively consistent with the correlation between natal-kick velocity and neutron-star birth mass predicted by \citet{spin2023}. Note that this is not true for the $1.9-2.0 \, M_\odot$ bin, and we discuss this at the end of this subsection. Also note that the removal fractions only become significantly large from the $1.5 - 1.6 \, M_\odot$ bin.

The combined black-hole and kick fractions produce a theoretical birth function that broadly reproduces the shape of the accretion-corrected observed distribution across the full mass range considered. The agreement is particularly strong between $1.3$ and $1.5\,M_\odot$ and again in the $1.7-2.0\,M_\odot$ bin, where the theoretical and observed probability densities are nearly identical. 

The largest low-mass residual occurs in the $1.2-1.3\,M_\odot$ bin, where the theoretical probability density exceeds the accretion-corrected observed value by approximately $27\%$. Therefore, some neutron stars formed in this mass range may be underrepresented in the measured sample because of binary evolution or observational selection effects. We discuss these possibilities further in \S\ref{issues}. Also, the existence of no theoretical birth function in the $1.1-1.2 \, M_\odot$ bin is discussed in \S\ref{issues}.

A residual enhancement remains in the $1.5-1.6\,M_\odot$ bin, where the theoretical mass functions overestimate the observed distribution by $\sim  34 \%$. This discrepancy is also seen in the birth-mass bin next to it ($1.6 - 1.7 M_\odot$), with an overestimate of $25.7\%$. These differences could be the result of our oversimple approach for accounting for the neutron star kick velocity correction. However, the major upshot of this is that, unlike in the case of accounting solely for black-hole fractions (which in itself did not explain the absence of the secondary peak in the accretion-corrected observed birth-mass function in the $1.5 - 1.7 M_\odot$ region), accounting for natal kicks actually removes this secondary enhancement in the theoretical birth function in this mass range. This reinforces the potential need to account for binary disruptions due to neutron-star kicks to understand the present-day observed mass distribution.

In the $1.8-1.9 M_\odot$ bin, our theoretical birth function overestimates the observed distribution by about $18\%$. This remaining excess may reflect the restricted black-hole masks adopted in this analysis. The calculations of \citet{ertl-2016} and \citet{mueller-2016}, for example, identify additional failed explosions near compactness maxima at $M_{\rm ZAMS}\simeq20.5-22\,M_\odot$. In the lower panel of Figure~\ref{fig:you+bh+kick}, progenitors associated with these compactness structures (progenitors in red starting from the compactness maximum at $\sim 20 M_\odot$ and extending to $22 M_\odot$ in dark orange) contribute primarily to the $1.8-1.9\,M_\odot$ birth-mass bin. Extending the black-hole-forming regions to include an appropriate fraction of these models would therefore selectively reduce the residual theoretical excess in precisely the mass interval where the present aggregate fit remains higher relative to the data.

Another limitation of this work is that after correcting for black hole formation and natal kicks, the theoretical birth function does not reproduce the highest mass tail ($2.0 - 2.2 \, M_\odot$) of the accretion-corrected observed mass distribution well.  We see no theoretical birth function in the mass range $2.0 - 2.2 \, M_\odot$, unlike the observed distribution. This discrepancy is because even the current state-of-the-art 3D core-collapse supernova (CCSN) simulation has yet to produce neutron stars with $M_{\mathrm{birth}} \ge 2.0\,  M_\odot$, where $M_{\mathrm{birth}}$ means the gravitational mass at birth~\citep{Burrows2025}. This is also why the $1.9-2.0\, M_\odot$ mass bin has a smaller $f_{\mathrm{kick}, i}$ than bins to its left. The CCSN simulations used to derive the compactness-to-birth-mass relation do not produce enough neutron stars in this mass range for the given set of progenitors, resulting in lower $f_{\mathrm{kick}, i}$, contradicting the theoretical predictions made in~\cite{spin2023}. Such limitations in CCSN simulations could result from the choice of progenitor models (see \S\ref{default}) and the equation of state (see \S\ref{issues}). Therefore, future studies are imperative with a varied choice of progenitor models and equations of state to resolve this highest-mass tail discrepancy.

The central result of this analysis is qualitative agreement in the overall shape between theory and observation. After accounting for black-hole formation, phenomenologically derived removal fractions motivated by binary unbinding effects in the theoretical birth function, and for post-birth accretion in the observed distribution, the two distributions exhibit the same broad mass-dependent structure, even though the probabilities assigned to individual mass bins differ. Both distributions show a sharp peak in the $1.2-1.3 \, M_\odot$ bin followed by a gradually fading tail as a function of increasing birth mass of neutron stars. The theoretical birth mass distribution reproduces the moderately higher mass tail in the $1.7 - 1.9 \, M_\odot$ bin reasonably well, and also shows some pileup of neutron stars in the $1.9 - 2.0 \, M_\odot$ bin. This supports the emerging claim that accretion alone is not the sole contributor to the heavier end of the observed distribution, and some neutron stars must be born massive~\citep{antoniadis, alsing, You2025}. It also points out that the three possible correction effects, namely accretion, black hole formation, and binary disruptions due to natal kicks, are important to sculpt the theoretically derived neutron star birth function in a way that agrees with the observation to some extent.

Finally, the aggregate fit provides the black hole fractions for progenitors in the $9-120 \, M_\odot$ ZAMS mass range to be $19.6\%$, and $80.4\%$ of the progenitors form neutron stars.
 
\subsection{Discussion of Outstanding Issues}
\label{issues}

This paper is meant to inaugurate the dialogue between the accumulating measurements of the neutron star mass distribution and detailed 3D core-collapse simulations.  The latter are beginning to reveal systematic correlations between the structure of progenitor stars, the birth mass of neutron stars, pulsar kick speeds, and black hole formation \citep{Burrows2024, Burrows2025}.
We find a concordance between theory and observations, but there remain many unknowns and topics unaddressed in this manuscript.  Foremost are the fact that we have not endeavored to make selection corrections to the observed neutron star populations, we have not addressed the potential metallicity dependence of neutron-star or black-hole birth, and we have not explored the important effect of binary evolution on the products of collapse. Concerning the latter, since most massive star progenitors are in multiplets, it is crucial that a distribution function of core carbon-oxygen (CO) mass for massive stars in the galaxy be determined, to replace the Salpeter function. In principle, with a CO core mass/compactness relationship and a distribution of CO masses we can recalculate a theoretical neutron-star birth mass function to reflect binary and multiple interactions. Moreover, we have leaned on the progenitor models of \citet{swbj16} and \citet{sukhbold2018}. The evolution of even single massive stars is not a converged theoretical problem, and one can anticipate revisions in the near future.

Furthermore, we note that our theoretical modeling does not explain neutron star masses below $\sim$1.1 $M_{\odot}$ and even disfavors neutron star masses much below $\sim$1.2 $M_{\odot}$. Many observational fits assume a Gaussian-like feature in this regime that tapers to these lower values by construction.

However, and importantly, if one assumes all stars in the 8$-$10 $M_{\odot}$ window supernova and leave neutron stars, there should be a pileup at a lower edge (see Figure \ref{fig:polyfit}), not a tapering. In fact, we do not find that these lowest-mass bins should be occupied; there should be a precipitous drop below $\sim1.15-1.2\, M_{\odot}$. The lowest neutron star gravitational mass found theoretically is 1.188 $M_{\odot}$ \citep{radice2017b},
with another study pegging this at 1.192 $M_{\odot}$ \citep{muller_minimum_2025}. These numbers are very weakly dependent on 1) the nuclear equation of state, which is still a vigorous subject of study \citep[e.g.,][]{2017ApJ...848..105T,2023Parti...6...30L,rusakov2026}, 2) the progenitor density structures, and 3) the vigor of supernova explosions at low compactness. So, the current theoretical limits are certainly not etched in granite. But, there doesn't seem to be too much latitude to decrease the lowest mass further (even for accretion-induced collapse), certainly below 1.1 $M_{\odot}$. Therefore, we suggest that any observed masses below 1.1 $M_{\odot}$ are due to the extension of mass error bars and are not the actual mass values. This is likely the case for the lowest-mass X-ray binaries with large error bars (\S\ref{measured}).

The three objects whose quoted central masses place them in the $1.1-1.2 \,M_\odot$ bin of the observed distribution
(Figure~\ref{fig:stacked_utd_hist}) are PSR J1741+1351
\citep{Arzoumanian2018}, the low-mass companion to PSR J0453+1559 \citep{Martinez2015}, and the secondary compact object in GW191219\_163120 \citep{Abbott2023}. However, the assignment of two of these objects to this bin is not robust. The neutron star in the PSR J1741+1351 NS-WD system has a measured mass of \(1.14^{+0.43}_{-0.25}\,M_\odot\), and its broad credible interval spans several histogram bins. Its placement in the \(1.1\text{-}1.2\,M_\odot\)
bin, therefore, could be a consequence of the uncertainty in the mass measurement. Similarly, the secondary in GW191219\_163120 has an inferred mass of \(1.17^{+0.07}_{-0.06}\,M_\odot\). Its \(90\%\) credible interval, \(1.11\text{-}1.24\,M_\odot\), has a total width of
\(0.13\,M_\odot\), exceeding the \(0.1\,M_\odot\) histogram bin width and extending into the \(1.2\text{-}1.3\,M_\odot\) bin. Thus, although its posterior median lies between \(1.1\) and \(1.2\,M_\odot\), the measurement does not uniquely localize the object to that interval. The only object with a mass posterior securely confined to the
\(1.1-1.2\,M_\odot\) bin is the low-mass companion to
PSR J0453+1559. Its mass, \(1.174\pm0.004\,M_\odot\), is sufficiently precise that its \(1\sigma\) interval lies entirely within the bin. Nevertheless, even the nature of this object is not completely settled: \citet{Tauris2019}
proposed that the companion may be a white dwarf rather than a neutron star. Consequently, the observational evidence for a genuine population of neutron stars in the \(1.1-1.2\,M_\odot\) range, and certainly in the 1.0$-$1.1 M$_{\odot}$ range, is weak. In contrast, the observed distribution exhibits a much more prominent
accumulation in the \(1.2-1.3\,M_\odot\) bin after accretion correction (Figure~\ref{fig:accretion}).

This dearth of securely identified neutron stars below
\(1.2\,M_\odot\) has important implications for both the minimum neutron-star birth mass and the minimum progenitor mass that successfully produces a neutron star. For a Salpeter initial mass function (IMF) with slope -2.35, progenitors between 8 and 9 $M_\odot$ account for approximately $15.1\%$ of total Salpeter IMF weight. If essentially all stars in this interval successfully explode and predominantly produce neutron stars in the $1.1-1.2 \,M_\odot$ range, their substantial IMF weight should generate a pronounced low-mass peak in the theoretical birth function. The absence of a corresponding feature in the observed distribution therefore disfavors the simultaneous assumptions that all $8-9\,M_\odot$ progenitors form neutron stars and that their remnants are concentrated below 1.2 $M_\odot$.

There are two broad ways to reconcile this discrepancy. First, the majority of $8-9 \,M_\odot$ progenitors may successfully explode, but produce neutron stars primarily in the $1.2-1.3 \,M_\odot$ interval. In this case, their IMF
contribution could help explain the observed pile-up in that bin, and the effective minimum neutron-star birth mass would be close to $\sim$1.2 $M_\odot$. This interpretation is qualitatively compatible with the default theoretical birth function in Figure~\ref{fig:birth-function-no-correction} and the birth function including only black-hole masking in Figure~\ref{fig:bh-mask}, for which the accretion-corrected observed distribution exceeds the theoretical prediction in the $1.2-1.3\,M_\odot$ bin. An additional contribution from low-mass progenitors could therefore improve the agreement in these cases. However, after both black-hole masking and natal-kick corrections are included (Figure~\ref{fig:you+bh+kick}), the theoretical birth function already exceeds the observed distribution in this mass range. Assigning the full $8-9\,M_\odot$ progenitor population to the $1.2-1.3\,M_\odot$ bin would then worsen the discrepancy. The reversal between these figures demonstrates that the inferred low-mass contribution remains sensitive to the adopted correction prescriptions and may expose oversimplifications in our present
implementation. However, our theoretical birth function would still successfully predict the minimum neutron star mass of near 1.2 $M_\odot$, despite these potential oversimplifications.

The second possibility is that only a fraction of the $8-9\,M_\odot$ population contributes to the neutron-star birth function, corresponding to an effective minimum progenitor mass greater than 8 $M_\odot$. This interpretation becomes particularly relevant because core-collapse simulations have produced remnants below 1.2 $M_\odot$ even from progenitors with initial masses of approximately 9 $M_\odot$ or greater
\citep{radice2017b,muller_minimum_2025}. It is therefore not evident that essentially all $8-9\,M_\odot$ progenitors should produce remnants exclusively in the $1.2-1.3\,M_\odot$ range. If such low-mass progenitors are capable of producing neutron stars below 1.2 $M_\odot$, but very few
such objects are present in the observed distribution, then a reduced explosion fraction in the $8-9\,M_\odot$ interval, or an effective lower progenitor-mass threshold for supernovae bigger than 8 $M_\odot$, becomes a natural explanation. A lower threshold of approximately 8.55 $M_\odot$, for example, has been suggested by \citet{quintana2026}. If the effective minimum mass for neutron-star formation lies even closer to 9 $M_\odot$, only a narrow portion of the $8-9\,M_\odot$ interval would contribute to the birth function. Its correspondingly small Salpeter weight could then produce only a few neutron stars in the $1.1-1.2 \,M_\odot$ bin, consistent with the current observational sample. Taken together, the scarcity of observed neutron stars below 1.2 $M_\odot$ motivates a reduced neutron-star formation efficiency between 8 and 9 $M_\odot$. The commonly adopted 8 $M_\odot$ lower limit should therefore be regarded as uncertain rather than as a firmly established boundary.

Finally, we note that the core-compactness to neutron-star birth mass mapping used in this work was derived using a suite of 3D simulations \citep{Burrows2024}, but for only the SFHo equation of state (EOS). Although the SFHo EOS is consistent with all current nuclear systematics and constraints \citep{2013ApJ...774...17S,richers:17}, exclusively relying on this model makes our mass function and its comparison with observations EOS-dependent. A majority of the previous literature exploring the EOS dependence of core-collapse supernovae was primarily concerned with explodability \citep{2013ApJ...764...10S, 2013ApJ...765...29C, burrows2018,PhysRevLett.124.092701,2026PhRvD.114b3003P}, and most did not evolve their models to late enough post-bounce times and for enough progenitors to provide a reliable constraint on the EOS dependence of the theoretical neutron-star mass function. The most recent study using the long-term 3D simulations with two different equations of state \citep{rusakov2026} considered only two progenitor models. Clearly, a comprehensive multi-progenitor set of 3D CCSN models with various equations of state, progenitor masses, and metallicities should be an important goal for future work.

\section{Conclusions}
\label{conclusion}

Supernova theory strongly suggests there should be a broad range in the masses of neutron stars at birth. The classical Chandrasekhar mass at 1.456 $M_{\odot}$ sets the overall scale, but pre-collapse stellar evolution and supernova dynamics together result in a wide spread in masses for neutron stars at birth.   

This wide spread of birth masses, both altered secondarily by subsequent accretion from a companion and sculpted by subtle masking to reflect black hole formation in islands of progenitor space, can explain the observed mass function of neutron stars in binaries. Importantly, black hole formation may not be monotonic with ZAMS mass $-$ there may be islands of black hole formation, some accompanied by supernova explosions \citep{Burrows2025}. Moreover, the breadth of the observed mass function is likely not solely (or even majorly) a reflection of differential accretion starting from a narrow birth function around ``$\sim$1.4 $M_{\odot}$."

The systematics of kick speeds with progenitor ZAMS mass and core compactness found in 3D core-collapse supernova simulations \citep{janka_kicks_2024,spin2023} should result in a difference between the neutron star birth mass function per se and that observed in surviving binaries. The lower predicted kick speeds expected during the birth of lower-mass neutron stars will select for such neutron stars in binaries that survive explosion. Along with this, the greater ``Salpeter" weight of their progenitors would further weight such a $\sim$1.4 $M_{\odot}$ peak, as indeed is seen in NS-NS systems (Figure \ref{fig:stacked_utd_hist}. In addition, the higher kick speeds expected at the birth of the higher mass neutron stars will increase the unbinding probability of those binaries, thereby shaving the observed mass function at the higher mass end above perhaps $\sim$1.6 $M_{\odot}$. 

The upshot of all these influences is to imprint upon the measured neutron star mass function features that reflect 
the varied physics of both neutron star origins and the neutron-star/black-hole dichotomy.  We emphasize that neither massive-star stellar evolution nor supernova theory are converged disciplines, merely quantitatively suggestive with modern insights garnered at great effort. Much work remains to be done in both sectors. Moreover, we have not attempted in this study to account for the numerous selection biases in the populations represented in Figure \ref{fig:stacked_utd_hist} and Table \ref{tab:unified_ns_masses}. Though it is likely the binary neutron star and Gaia samples have the least bias \citep{2026ApJ...998L..14S}, the current collection of neutron stars with measured masses is quite heterogeneous. Some examples (not exhaustive) of potential biases include 1) for spectroscopic binaries, biases in inclination \citep{2010MNRAS.409..199F}; 2) in general, Malmquist bias \citep{1992ApJ...391..494L}; 3) for X-ray sources, biases towards the brighter cohort \citep{2002A&A...391..923G}; and 4) for radio pulsars, selecting for higher luminosities \citep{faucher_kaspi} and not correcting for direction-dependent dispersion measures and galactic electron density distributions \citep{2017ApJ...835...29Y, 2002astro.ph..7156C} Finally, we have not addressed the potential effects of stellar metallicity, nor pre-explosion binarity. Massive stars are overwhelmingly found in multiple star systems. 

In fact, given the prevalence of multiplicity in the population of massive star supernova progenitors and their relatively compact orbits (increasing the chance of mass transfer), instead of a Salpeter (or equivalent) distribution of ZAMS mass, one may actually need a distribution of carbon core masses \citep{maltsev2025}, and a corresponding carbon/oxygen (CO)-core-mass/compactness relation. Thus, substituting for the mapping depicted in Figure \ref{fig:eta} and the Salpeter distribution would be a population-synthesis-derived CO-core mass distribution function, with its associated compactness/CO-core mapping. With these we could then use our compactness/$M_{\mathrm{birth}}$ mapping (e.g., Figure \ref{fig:polyfit} and Table 1) to derive the birth neutron star mass function. This desired synthesis has yet to emerge, but a community is in active pursuit \citep{laplace2021,schneider2024,schneider2025_interacting,ercolino2026}.

Nevertheless, our analysis illuminates the emerging dialogue between the observation of compact objects and modern supernova theory that addresses in some detail their birth. We suggest that supernova theory can now be used to explain, however provisionally, various measured attributes of the population of compact objects and that an era of productive engagement between detailed supernova theory and observation is at hand.

\section*{Data Availability}  

The data presented in the figures and in the tables in the Appendix can be found online in our zenodo upload: \hyperlink{https://zenodo.org/records/23018776}{https://zenodo.org/records/23018776}.

\section*{Acknowledgments}

We thank Tianshu Wang for his significant contributions over the years to our supernova simulation efforts. DV acknowledges support from the NASA Hubble Fellowship Program grant HST-HF2-51520. AB acknowledges former support from the U.~S.\ Department of Energy Office of Science and the Office of Advanced Scientific Computing Research via the Scientific Discovery through Advanced Computing (SciDAC4) program and Grant DE-SC0018297 (subaward 00009650) and former support from the U.~S.\ National Science Foundation (NSF) under Grant AST-1714267. We are happy to acknowledge access to the Frontera cluster (under awards AST20020 and AST21003). This research is part of the Frontera computing project at the Texas Advanced Computing Center \citep{Stanzione2020}. Frontera is made possible by NSF award OAC-1818253. Additionally, a generous award of computer time was provided by the INCITE program, enabling this research to use resources of the Argonne Leadership Computing Facility, a DOE Office of Science User Facility supported under Contract DE-AC02-06CH11357. Finally, the authors acknowledge computational resources provided by the high-performance computer center at Princeton University, which is jointly supported by the Princeton Institute for Computational Science and Engineering (PICSciE) and the Princeton University Office of Information Technology, and our continuing allocation at the National Energy Research Scientific Computing Center (NERSC), which is supported by the Office of Science of the U.~S.\ Department of Energy under contract DE-AC03-76SF00098. W.K. acknowledges the use of \texttt{Grammarly} writing tool to improve spelling, sentence structure, and word choice in the manuscript writing process.

\bibliographystyle{aasjournal}
\bibliography{References}


\begin{deluxetable*}{ccc}
\tablecolumns{3}
\tablewidth{100pt}
\label{table1_new}
\begin{minipage}{\textwidth}
   \centering
   \textbf{Table 1: 3D Model Suite and Parameters} \\
\end{minipage}

\tablehead{
\colhead{Progenitor} & 
\colhead{$\xi$(1.75)} & 
\colhead{Gravitational Mass ($M_{\odot}$)}
}
\startdata
9a              & 6.7$\times 10^{-5}$ & 1.237 \\
9b              & 6.7$\times 10^{-5}$ & 1.238 \\
9.25            & 2.5$\times 10^{-3}$ & 1.263 \\
9.5             & 8.5$\times 10^{-3}$ & 1.278 \\
11              & 0.12                & 1.361 \\
{\bf 12.25}     & {\bf 0.34}          & $\sim${\bf  11.1} \\
13.63           & 0.486               & 1.609 \\
{\bf 14}        & {\bf 0.48}          & $\sim${\bf  12.1} \\
15.01           & 0.29                & 1.474 \\
16              & 0.35                & 1.505 \\
16.5            & 0.568               & 1.648 \\
17              & 0.74                & 1.785 \\
17.5            & 0.796               & 1.870 \\
18              & 0.37                & 1.510 \\
18.5            & 0.80                & 1.854 \\
19              & 0.48                & 1.603 \\
19.5            & 0.649               & 1.711 \\
{\bf 19.56}     & {\bf 0.85}          & $\sim${\bf 3.0} \\
20              & 0.79                & 1.857 \\
21.68           & 0.84                & 1.882 \\
{\bf 23}        & 0.74                & $\sim${\bf 5.0} \\
24              & 0.77                & 1.773 \\
25              & 0.80                & 1.830 \\
35              & 0.855               & 1.945 \\
{\bf 40}        & {\bf 0.884}         & {\bf $\sim$9.0} \\
60              & 0.440               & 1.588 \\
{\bf 100}       & {\bf 1.02, 0.81}    & $\sim${\bf  37} \\
\hline
\enddata
\caption{3D Model mappings. This table contains the mapping between progenitor mass, compactness, and derived neutron star birth gravitational masses from \citet{Burrows2024}. The first column contains the ZAMs mass value of the candidate progenitor. The units of the first column are in \(M_\odot\). The second column lists compactness $\xi(1.75)$, and the third column lists the gravitational mass of the resulting neutron star found in the 3D simulation. Bold entries denote BH systems; for those rows, the reported final mass of the black hole $M_{\rm final}$ has been inserted in place of the gravitational mass of the neutron star $M_{\rm grav}$ \citep{Burrows2025}. Most data values come directly from Table 1 of \citep{Burrows2024}. The remaining compactness and gravitational birth mass values for progenitor masses 13.63, 16.5, 17.5, 21.68, 35, and 100 come from other 3D {\sc{Fornax}} simulations.}
\end{deluxetable*}

\onecolumngrid

\clearpage
\setlength{\tabcolsep}{0pt}
\startlongtable

\begin{deluxetable*}{@{}ccccccc}
\tabletypesize{\normalsize}
\tablecolumns{7}
\tablewidth{100pt}
\tablecaption{\bf{{\normalsize{Observed Neutron Stars}}}}
\label{tab:unified_ns_masses}
\tablehead{
\colhead{Name} &
\colhead{Type} &
\colhead{$m_p$ ($M_\odot$)} &
\colhead{$m_c$ ($M_\odot$)} &
\colhead{$P$ (ms)} &
\colhead{$e_0$} &
\colhead{Reference}
}
\startdata
\cutinhead{Double neutron star systems}
J0514$-$4002A & NS--NS & 1.25(6) & 1.22(6) & 4.99 & 0.888 & \citep{Ridolfi2019} \\
J1946$+$2052 & NS--NS & 1.2842(21) & 1.2477(21) & 16.96 & 0.06384 & \citep{Meng2025} \\
J1829$+$2456 & NS--NS & 1.306(4) & 1.299(4) & 41.01 & 0.139 & \citep{Haniewicz2021} \\
J0641$+$0448 & NS--NS & $1.319^{+0.021}_{-0.035}$ & $1.269^{+0.022}_{-0.016}$ & 25.04 & 0.14545 & \citep{Yang2026} \\
J1906$+$0746 & NS--NS & 1.322(11) & 1.291(11) & 144.1 & 0.085 & \citep{vanLeeuwen2015} \\
B1534$+$12 & NS--NS & 1.3330(2) & 1.3455(2) & 37.90 & 0.274 & \citep{Fonseca2014} \\
J0737$-$3039A & NS--NS & 1.3381(7) & 1.2489(7) & 22.70 & 0.088 & \citep{Kramer2006} \\
J0509$+$3801 & NS--NS & 1.34(8) & 1.46(8) & 76.54 & 0.586 & \citep{Lynch2018} \\
J1757$-$1854 & NS--NS & 1.3406(5) & 1.3922(5) & 21.50 & 0.606 & \citep{Cameron2023} \\
J1756$-$2251 & NS--NS & 1.341(7) & 1.230(7) & 28.46 & 0.181 & \citep{Ferdman2014} \\
B2127$+$11C & NS--NS & 1.358(10) & 1.354(10) & 30.53 & 0.681 & \citep{Jacoby2006} \\
J1807$-$2500B & NS--NS & 1.3655(21) & 1.2064(21) & 4.19 & 0.747 & \citep{Lynch2012} \\
B1913$+$16 & NS--NS & 1.4398(2) & 1.3886(2) & 59.03 & 0.617 & \citep{Weisberg2010} \\
J1518$+$4904 & NS--NS & $1.470^{+0.030}_{-0.034}$ & $1.248^{+0.035}_{-0.029}$ & 40.94 & 0.24948 & \citep{Tan2024} \\
J0453$+$1559 & NS--NS & 1.559(5) & 1.174(4) & 45.78 & 0.113 & \citep{Martinez2015} \\
J1913$+$1102 & NS--NS & 1.62(3) & 1.27(3) & 27.29 & 0.090 & \citep{Ferdman2020} 
\\
\cutinhead{Neutron star--white dwarf systems}
J1802$-$2124 & NS--WD & 1.24(11) & 0.78 & 12.65 & $10^{-6}$ & \citep{Ferdman2010} \\
J2045$+$3633 & NS--WD & 1.251(21) & 0.873 & 31.68 & 0.017 & \citep{McKee2020} \\
J1141$-$6545 & NS--WD & 1.27(1) & 1.01 & 393.9 & 0.172 & \citep{VenkatramanKrishnan2020} \\
J2023$+$2853 & NS--WD & 1.28(6) & 0.85(2) & 11.33 & $1.31\times10^{-5}$ & \citep{Yang2025} \\
J1918$-$0642 & NS--WD & 1.29(10) & 0.231 & 7.646 & $10^{-5}$ & \citep{Arzoumanian2018} \\
J1910$-$5959A & NS--WD & 1.33(11) & 0.18 & 3.266 & $10^{-6}$ & \citep{Corongiu2012} \\
J1949$+$3106 & NS--WD & 1.34(17) & 0.81 & 13.14 & $10^{-5}$ & \citep{Zhu2019} \\
J1543$-$5149 & NS--WD & $1.349^{+0.043}_{-0.061}$ & $0.2233^{+0.0045}_{-0.0061}$ & 2.057 & $2.13\times10^{-5}$ & \citep{Chisabi2025} \\
J1713$+$0747 & NS--WD & 1.35(7) & 0.292 & 4.57 & $10^{-4}$ & \citep{Arzoumanian2018} \\
J2234$+$0611 & NS--WD & 1.353(17) & 0.298 & 3.577 & 0.129 & \citep{Stovall2019} \\
J2305$+$4707 & NS--WD & 1.38(10) & 1.26 & 1066 & 0.658 & \citep{Thorsett1999} \\
J2043$+$1711 & NS--WD & 1.38(13) & 0.173 & 2.38 & $10^{-6}$ & \citep{Arzoumanian2018} \\
J2053$+$4650 & NS--WD & 1.40(21) & 0.86 & 12.59 & $10^{-5}$ & \citep{Berezina2017} \\
J1933$-$6211 & NS--WD & $1.4^{+0.3}_{-0.2}$ & 0.43(5) & 3.543 & $1.26\times10^{-6}$ & \citep{Geyer2023} \\
J0337$+$1715 & NS--WD & 1.4359(3) & 0.1973 & 2.733 & $10^{-3}$ & \citep{Archibald2018} \\
J0437$-$4715 & NS--WD & 1.44(7) & 0.224 & 5.76 & $10^{-5}$ & \citep{Reardon2016} \\
J1012$-$4235 & NS--WD & $1.44^{+0.13}_{-0.12}$ & $0.270^{+0.016}_{-0.015}$ & 3.101 & 0.000346 & \citep{Gautam2024} \\
J1738$+$0333 & NS--WD & 1.47(7) & 0.181 & 5.85 & $10^{-7}$ & \citep{Freire2012} \\
J1909$-$3744 & NS--WD & 1.486(11) & 0.208 & 2.95 & $10^{-7}$ & \citep{Reardon2021} \\
J0218$+$4232 & NS--WD & $1.49^{+0.23}_{-0.20}$ & $0.179^{+0.018}_{-0.016}$ & 2.323 & $6.3\times10^{-6}$ & \citep{Tan2024} \\
J1950$+$2414 & NS--WD & 1.496(23) & 0.28 & 4.30 & 0.08 & \citep{Zhu2019} \\
J1125$-$6014 & NS--WD & 1.5(2) & 0.31 & 2.63 & $10^{-6}$ & \citep{Reardon2021} \\
J1857$+$0943 & NS--WD & 1.54(13) & 0.263 & 5.36 & $10^{-5}$ & \citep{Reardon2021} \\
J1227$-$6208 & NS--WD & 1.54(15) & 1.40(7) & 34.53 & 0.00115 & \citep{ColomiBernadich2024} \\
J0751$+$1807 & NS--WD & 1.64(15) & 0.16 & 3.48 & $10^{-6}$ & \citep{Desvignes2016} \\
J1748$-$2446ap & NS--WD & $1.700^{+0.015}_{-0.045}$ & $0.294^{+0.046}_{-0.014}$ & 3.745 & 0.90519 & \citep{Padmanabh2024} \\
J0955$-$6150 & NS--WD & 1.71(3) & 0.254 & 2.0 & 0.118 & \citep{Serylak2022} \\
J1012$+$5307 & NS--WD & 1.72(16) & 0.165 & 5.26 & $10^{-6}$ & \citep{Mata2020} \\
J1600$-$3053 & NS--WD & 1.77(36) & 0.34 & 3.598 & $10^{-3}$ & \citep{Reardon2021} \\
J1946$+$3417 & NS--WD & 1.828(22) & 0.2556 & 3.17 & 0.134 & \citep{Barr2017} \\
J2222$-$0137 & NS--WD & 1.831(10) & 1.319 & 32.82 & $10^{-3}$ & \citep{Guo2021} \\
J1943$+$2210 & NS--WD & $1.84^{+0.11}_{-0.09}$ & $1.03^{+0.04}_{-0.03}$ & 12.87 & $1.6\times10^{-6}$ & \citep{Yang2025} \\
J1614$-$2230 & NS--WD & 1.908(16) & 0.493 & 3.15 & $10^{-6}$ & \citep{Arzoumanian2018} \\
J0348$+$0432 & NS--WD & 2.01(4) & 0.172 & 39.12 & $10^{-6}$ & \citep{Antoniadis2013} \\
J0740$+$6620 & NS--WD & 2.08(7) & 0.253 & 2.89 & $10^{-6}$ & \citep{Fonseca2021} \\
J0621$+$1002 & NS--WD & $1.53^{+0.10}_{-0.20}$ & -- & -- & -- & \citep{Kasian2012} \\
J1017$-$7156 & NS--WD & 2.0$\pm$0.8 & -- & -- & -- & \citep{Reardon2021} \\
J1022$+$1001 & NS--WD & 1.44$\pm$0.44 & -- & -- & -- & \citep{Reardon2021} \\
B1516$+$02B & NS--WD & 2.08$\pm$0.19 & -- & -- & -- & \citep{Freire2008} \\
J1528$-$3146 & NS--WD & $1.61^{+0.14}_{-0.13}$ & -- & -- & -- & \citep{Berthereau2023} \\
J1741$+$1351 & NS--WD & $1.14^{+0.43}_{-0.25}$ & -- & -- & -- & \citep{Arzoumanian2018} \\
J1748$-$2446am & NS--WD & $1.649^{+0.037}_{-0.11}$ & -- & -- & -- & \citep{Bridget2018} \\
B1802$-$07 & NS--WD & $1.26^{+0.08}_{-0.17}$ & -- & -- & -- & \citep{Thorsett1999} \\
J1811$-$2405 & NS--WD & $2.0^{+0.8}_{-0.5}$ & -- & -- & -- & \citep{Ng2020} \\
\cutinhead{Neutron star plus main--sequence star systems}
LAMOST J1123 & NS--MS & 1.24(3) & 0.61 & -- & -- & \citep{Yi2022} \\
PSR J0045$-$7319 & NS--MS & 1.58(34) & 10 & 926.3 & -- & \citep{Arzoumanian2018} \\
PSR J1903$+$0327 & NS--MS & 1.666(12) & 1.033 & 2.15 & -- & \citep{Thorsett1999} \\
\cutinhead{Gaia-detected systems}
J0553$-$1349 & NS--MS & $1.33 \pm 0.05$ & $0.98 \pm 0.06$ & -- & $0.3879 \pm 0.0007$ & \citep{2014ApJ...786...83T} \\
J2057$-$4742 & NS--MS & $1.31 \pm 0.04$ & $1.048 \pm 0.031$ & -- & $0.3095 \pm 0.0026$ & \citep{Badry2024Gaia} \\
J1553$-$6846 & NS--MS & $1.323 \pm 0.032$ & $1.04 \pm 0.05$ & -- & $0.5314 \pm 0.0021$ & \citep{Badry2024Gaia} \\
J2102$+$3703 & NS--MS & $1.473 \pm 0.034$ & $1.03 \pm 0.03$ & -- & $0.448 \pm 0.009$ & \citep{Badry2024Gaia} \\
J0742$-$4749 & NS--MS & $1.28 \pm 0.04$ & $0.90 \pm 0.05$ & -- & $0.168 \pm 0.004$ & \citep{Badry2024Gaia} \\
J0152$-$2049 & NS--MS & $1.291 \pm 0.024$ & $0.782 \pm 0.03$ & -- & $0.6615 \pm 0.0010$ & \citep{Badry2024Gaia} \\
J0003$-$5604 & NS--MS & $1.34 \pm 0.04$ & $0.802 \pm 0.03$ & -- & $0.795 \pm 0.005$ & \citep{Badry2024Gaia} \\
J1733$+$5808 & NS--MS & $1.362 \pm 0.030$ & $1.16 \pm 0.05$ & -- & $0.3093 \pm 0.0010$ & \citep{Badry2024Gaia} \\
J1150$-$2203 & NS--MS & $1.39 \pm 0.04$ & $1.18 \pm 0.06$ & -- & $0.552 \pm 0.004$ & \citep{Badry2024Gaia} \\
J1449$+$6919 & NS--MS & $1.258 \pm 0.032$ & $0.91 \pm 0.05$ & -- & $0.2668 \pm 0.0010$ & \citep{Badry2024Gaia} \\
J0217$-$7541 & NS--MS & $1.396 \pm 0.033$ & $0.996 \pm 0.033$ & -- & $0.3228 \pm 0.0033$ & \citep{Badry2024Gaia} \\
J0639$-$3655 & NS--MS & $1.70 \pm 0.07$ & $1.32 \pm 0.06$ & -- & $0.721 \pm 0.013$ & \citep{Badry2024Gaia} \\
J1739$+$4502 & NS--MS & $1.38 \pm 0.04$ & $0.781 \pm 0.03$ & -- & $0.6777 \pm 0.0018$ & \citep{Badry2024Gaia} \\
J0036$-$0932 & NS--MS & $1.362 \pm 0.034$ & $0.94 \pm 0.04$ & -- & $0.3993 \pm 0.0021$ & \citep{Badry2024Gaia} \\
J1432$-$1021 & NS--MS & $1.898 \pm 0.030$ & $0.79 \pm 0.03$ & -- & $0.1203 \pm 0.0022$ & \citep{Badry2024Gaia} \\
J1048$+$6547 & NS--MS & $1.52 \pm 0.07$ & $0.99 \pm 0.05$ & -- & $0.357 \pm 0.009$ & \citep{Badry2024Gaia} \\
J2145$+$2837 & NS--MS & $1.396 \pm 0.035$ & $0.95 \pm 0.05$ & -- & $0.5840 \pm 0.0035$ & \citep{Badry2024Gaia} \\
J2244$-$2236 & NS--MS & $1.443 \pm 0.023$ & $1.002 \pm 0.03$ & -- & $0.5666 \pm 0.0011$ & \citep{Badry2024Gaia} \\
J0824$+$5254 & NS--MS & $1.604 \pm 0.034$ & $1.102 \pm 0.03$ & -- & $0.686 \pm 0.012$ & \citep{Badry2024Gaia} \\
J0230$+$5950 & NS--MS & $1.401 \pm 0.034$ & $1.114 \pm 0.03$ & -- & $0.753 \pm 0.011$ & \citep{Badry2024Gaia} \\
J0634$+$6256 & NS--MS & $1.48 \pm 0.09$ & $1.18 \pm 0.06$ & -- & $0.564 \pm 0.011$ & \citep{Badry2024Gaia} \\
\cutinhead{Redback binary pulsar systems}
PSR J1723$-$2837 & RB & 1.22(26) & 0.36 & 1.856 & -- & \citep{Strader2019} \\
PSR J2039$-$5617 & RB & 1.3(1) & 0.2 & 2.651 & -- & \citep{Clark2021} \\
PSR J1023$+$0038 & RB & 1.71(16) & 0.24 & 1.688 & -- & \citep{Deller2012} \\
PSR J2129$-$0429 & RB & 1.74(18) & 0.44 & 7.62 & -- & \citep{Bellm2016} \\
PSR J2215$+$5135 & RB & 2.28(10) & 0.25 & 2.610 & -- & \citep{Kandel2020} \\
3FGL J0212.1$+$5320 & RB & $1.85^{+0.32}_{-0.26}$ & -- & -- & -- & \citep{Shahbaz2017} \\
3FGL J0427.9$-$6704 & RB & $1.86^{+0.11}_{-0.10}$ & -- & -- & -- & \citep{Strader2016} \\
2FGL J0846.0$+$2820 & RB & 1.96$\pm$0.41 & -- & -- & -- & \citep{Swihart2017} \\
1FGL J1417.7$-$4407 & RB & $1.62^{+0.43}_{-0.17}$ & -- & -- & -- & \citep{Swihart2018} \\
\cutinhead{Black widow binary pulsar systems}
PSR J2339$-$0533 & BW & 1.64(27) & 0.35 & 2.884 & -- & \citep{Strader2019} \\
PSR J1555$-$2908 & BW & 1.67(7) & 0.06 & 1.788 & -- & \citep{Kennedy2022} \\
PSR J1301$+$0833 & BW & 1.74(20) & 0.035 & 1.84 & -- & \citep{Romani2016} \\
PSR J1810$+$1744 & BW & 2.13(4) & 0.065 & 1.66 & -- & \citep{Romani2021} \\
PSR J1653$-$0158 & BW & 2.17(21) & 0.014 & 1.968 & -- & \citep{Nieder2020} \\
PSR J1959$+$2048 & BW & 2.18(9) & 0.033 & 1.607 & -- & \citep{Kandel2020} \\
PSR J0952$-$0607 & BW & 2.35(17) & 0.032 & 1.41 & -- & \citep{Romani2022} \\
J1311$-$3430 & BW & 2.22$\pm$0.1 & -- & -- & -- & \citep{Kandel2023} \\
\cutinhead{Isolated Neutron Stars}
PSR J0030$+$0451 & INS & 1.44(15) & -- & 4.865 & -- & \citep{Miller2019} \\
\cutinhead{Low-mass X-ray binaries}
4U 1608$-$52 & LMXB & 1.57(30) & -- & 1.61 & -- & \citep{Ozel2016} \\
KS 1731$-$260 & LMXB & 1.61(37) & -- & 1.91 & -- & \citep{Ozel2016} \\
X1822$-$371 & LMXB & 1.69(13) & 0.46 & 590 & -- & \citep{Iaria2015} \\
2S 0921$-$630 & LMXB & 1.44$\pm$0.1 & -- & -- & -- & \citep{Steeghs2007} \\
4U1702$-$429 & LMXB & 1.9$\pm$0.3 & -- & -- & -- & \citep{Nattila2017} \\
4U 1724$-$207 & LMXB & $1.81^{+0.25}_{-0.37}$ & -- & -- & -- & \citep{Ozel2016} \\
4U 1820$-$30 & LMXB & $1.77^{+0.25}_{-0.28}$ & -- & -- & -- & \citep{Ozel2016} \\
Cyg X-2 & LMXB & 1.71$\pm$0.21 & -- & -- & -- & \citep{Casares2010} \\
EXO 1745$-$248 & LMXB & $1.65^{+0.21}_{-0.31}$ & -- & -- & -- & \citep{Ozel2016} \\
SAX J1748.9$-$2021 & LMXB & $1.81^{+0.25}_{-0.37}$ & -- & -- & -- & \citep{Ozel2016} \\
XTE J2123$-$058 & LMXB & 1.53$\pm$0.42 & -- & -- & -- & \citep{Gelino2002} \\
\cutinhead{High-mass X-ray binaries}
4U 1538$-$522 & HMXB & 1.02(17) & 16 & 526.8 & -- & \citep{Iaria2015} \\
SMC X$-$1 & HMXB & 1.21(12) & 18 & 0.71 & -- & \citep{Falanga2015} \\
XTE J1855$-$026 & HMXB & 1.41(24) & 21 & 360.7 & -- & \citep{Falanga2015} \\
LMC X$-$4 & HMXB & 1.57(11) & 18 & 13.5 & -- & \citep{Falanga2015} \\
Cen X$-$3 & HMXB & 1.57(16) & 24 & 4.8 & -- & \citep{Falanga2015} \\
SAX J1802.7$-$2017 & HMXB & 1.57(25) & 22 & 139.6 & -- & \citep{Falanga2015} \\
OAO 1657$-$415 & HMXB & 1.74(30) & 17.5 & 37.3 & -- & \citep{Falanga2015} \\
EXO 1722$-$363 & HMXB & 1.91(45) & 18 & 413.9 & -- & \citep{Falanga2015} \\
4U 1700$-$377 & HMXB & 1.96(19) & 46 & -- & -- & \citep{Falanga2015} \\
J013236.7$+$303228 & HMXB & 2.0(4) & 11 & -- & -- & \citep{Bhalerao2012} \\
Vela X$-$1 & HMXB & 2.12(16) & 26 & 283.2 & -- & \citep{Falanga2015} \\
Her X-1 & HMXB & 1.07$\pm$0.36 & -- & -- & -- & \citep{Rawls2011} \\
\cutinhead{Binary neutron star mergers}
GW170817 & NS--NS & $1.34^{+0.12}_{-0.09}$ & $1.38^{+0.11}_{-0.11}$ & -- & -- & \citep{Abbott2017} \\
GW190425 & NS--NS & $1.64^{+0.13}_{-0.11}$ & $1.66^{+0.12}_{-0.12}$ & -- & -- & \citep{Abbot2020} \\
\cutinhead{Neutron star--black hole mergers}
GW191219 & NS--BH & $1.17^{+0.07}_{-0.06}$ & 31.1 & -- & -- & \citep{Abbott2023} \\
GW200115 & NS--BH & $1.4^{+0.6}_{-0.2}$ & 5.9 & -- & -- & \citep{Abbott2021} \\
GW200105 & NS--BH & $1.9^{+0.2}_{-0.2}$ & 8.9 & -- & -- & \citep{Abbott2021} \\
\enddata
\tablecomments{Mass table combining the double neutron star, neutron star--white dwarf, neutron star plus main-sequence star, Gaia-detected systems, redback and black widow pulsars, isolated neutron stars, X-ray binaries, and compact-object merger events. This includes neutron-star/system name, system type, neutron star mass ($m_p$), companion-star mass ($m_c$), spin period ($P$), eccentricity ($e_0$), and a reference column. Gaia-detected systems are labeled separately by listing the unseen compact-object mass under $m_p$, the luminous companion under $m_c$, $P$, and $e_0$, equal to the fitted eccentricity. Dashes indicate unavailable or inapplicable parameters. NS--NS = double neutron star; NS--WD = neutron star--white dwarf; NS--MS = neutron star plus main-sequence star; RB = redback; BW = black widow; INS = isolated neutron star; LMXB/HMXB = low-/high-mass X-ray binary. Given rapid changes in the field, this table should not be considered complete. \\}
\end{deluxetable*}


\section*{}

\label{lastpage}

\end{document}